\documentclass[a4paper, twocolumn]{article}
\usepackage{palatino} 
\usepackage{graphicx} 
\usepackage{amsmath} 
\usepackage{amssymb}
\usepackage{hyperref} 
\usepackage{geometry} 
\usepackage{changepage}
\usepackage{IEEEtrantools} 
\usepackage{booktabs} 
\usepackage{multirow}
\usepackage{subcaption}
\usepackage{color}
\usepackage[dvipsnames]{xcolor}
\usepackage{bm}
\usepackage{orcidlink}

\hypersetup{
    colorlinks=true,        
    linkcolor=blue,        
    citecolor=green,       
    urlcolor=blue           
}

\title{\fontsize{12.5}{15}\selectfont{\textbf{CQ-CNN: A Hybrid Classical-Quantum Convolutional Neural Network for Alzheimer’s Disease Detection Using Diffusion-Generated and U-Net Segmented 3D MRI}}}
\author{
    \vspace{0.3em}\text{Mominul Islam}$^{1, \dagger}\hspace{0.1em}\orcidlink{0009-0001-6409-964X}$
    \hspace{1.0em}\text{Mohammad Junayed Hasan}$^{2, \dagger}\hspace{0.1em}\orcidlink{0009-0008-3451-0267}$
    \hspace{1.0em}\text{M.R.C. Mahdy}$^{1}$\thanks{Corresponding author \\ \textit{Email addresses:} \href{mailto:mominul.islam05@northsouth.edu}{mominul.islam05@northsouth.edu} (M. Islam), \\ \href{mailto:mhasan21@jhu.edu}{mhasan21@jhu.edu} (M. Hasan), \\
    \href{mailto:mahdy.chowdhury@northsouth.edu}{mahdy.chowdhury@northsouth.edu} (M.R.C. Mahdy)}\hspace{0.4em}$\orcidlink{0000-0003-3737-0315}$ \\
    $^1$\normalsize\textit{Department of Electrical and Computer Engineering, North South University}\\
    $^2$\normalsize\textit{Department of Computer Science, Johns Hopkins University}\\
    $^{\dagger}$\normalsize\textit{Mahdy Research Academy}
}

\date{}
\begin{document}

\maketitle

\begin{center}
\section*{Abstract}
\end{center}

\begin{adjustwidth}{0.45cm}{0.45cm}
\small
The detection of Alzheimer’s disease (AD) from clinical MRI data is an active area of research in medical imaging. Recent advances in quantum computing, particularly the integration of parameterized quantum circuits (PQCs) with classical machine learning architectures, offer new opportunities to develop models that may outperform traditional methods. However, quantum machine learning (QML) remains in its early stages and requires further experimental analysis to better understand its behavior and limitations. In this paper, we propose an end-to-end hybrid classical-quantum convolutional neural network (CQ-CNN) for AD detection using clinically formatted 3D MRI data. Our approach involves developing a framework to make 3D MRI data usable for machine learning, designing and training a brain tissue segmentation model (SkullNet), and training a diffusion model to generate synthetic images for the minority class. Our converged models exhibit potential quantum advantages, achieving higher accuracy in fewer epochs than classical models. The proposed \(\beta_8\)-3-qubit model achieves an accuracy of 97.50\%, surpassing state-of-the-art (SOTA) models while requiring significantly fewer computational resources. In particular, the architecture employs only 13K parameters (0.48 MB), reducing the parameter count by more than 99.99\% compared to current SOTA models. Furthermore, the diffusion-generated data used to train our quantum models, in conjunction with real samples, preserve clinical structural standards, representing a notable first in the field of QML. We conclude that CQ-CNN architecture-like models, with further improvements in gradient optimization techniques, could become a viable option and even a potential alternative to classical models for AD detection, especially in data-limited and resource-constrained clinical settings.
\end{adjustwidth}

\vspace{0.25cm}
\noindent \textit{\textbf{Keywords:}} Alzheimer's disease detection, Brain tissue segmentation, U-Net, Probabilistic diffusion model, Quantum neural network, Qiskit.

\section{Introduction}
Alzheimer's disease (AD) is a progressive neurodegenerative disorder that mainly affects the elderly, leading to cognitive decline and memory loss \cite{pichet2024proteomic, pourhadi2024restoring}. It is the most common form of dementia and poses a growing challenge to healthcare systems worldwide, with more than 55 million people living with the condition \cite{sweidan2024explainability}. By 2050, it is estimated that one in every 85 people will be diagnosed with AD \cite{orouskhani2022alzheimer}. Currently, there is no proven cure or way to reverse the progression of AD \cite{przybyszewski2024cure, elazab2024alzheimer}. AD is primarily managed through supportive care provided by healthcare professionals \cite{bhandarkar2024deep}. AD pathology is characterized by the accumulation of abnormal proteins, such as amyloid $\beta$ ($A\beta$) and tau ($\tau$), in the brain. These proteins interfere with the communication between brain cells, altering their function and ultimately leading to cell death \cite{thal2022central, zhang2021interaction}. As brain cells die, key areas involved in cognition, particularly the hippocampus, begin to shrink. The hippocampus plays a crucial role in memory formation and retrieval, and its degeneration is closely related to the memory loss characteristic of AD \cite{thompson2004mapping, llorens2014selective}.

Structural changes in the brain can be detected using imaging techniques such as positron emission tomography (PET), cerebrospinal fluid (CSF) analysis, and magnetic resonance imaging (MRI) \cite{palmqvist2015detailed, de2017cerebrospinal}. PET scans use radioactive tracers to highlight areas of the brain with abnormal metabolic activity and require the injection of a radioactive substance \cite{toyama2005pet, buther2013external}. CSF analysis often involves a lumbar puncture, a procedure in which a needle is inserted into the lower back to collect fluid surrounding the spinal cord. This fluid can reveal important biomarkers for AD, such as the levels of $A\beta$ and $\tau$ proteins \cite{wright2012cerebrospinal, gordon2016relationship}. Both PET and CSF are invasive methods. In contrast, MRI is a non-invasive imaging technique that does not require injections, radiation, or other procedures that penetrate the body. MRI uses powerful magnetic fields and radio waves to create highly detailed images of brain structure, allowing the identification of physical changes such as brain atrophy or shrinkage \cite{probert1999magnetic, viola2015towards}.

However, manually interpreting MRI scans is time-consuming and requires expert knowledge \cite{vemuri2010role}. This has led to a push to develop automated diagnostic systems that can quickly and accurately analyze brain images. In recent years, machine learning, especially convolutional neural networks (CNNs), has shown great promise in AD detection by automatically extracting features from MRI images \cite{li2019deep}. Researchers have also combined traditional methods such as Support Vector Machines (SVM) and Random Forests with feature extraction techniques such as wavelet entropy and principal component analysis (PCA) to further improve model performance \cite{kloppel2008automatic, gray2013random}.

Despite these advances, there are still challenges in accurately diagnosing AD, particularly in its early stages, when structural changes in the brain are less pronounced. Moreover, the lack of large and diverse datasets, along with issues such as class imbalance, can limit the effectiveness of these models \cite{islam2024cossif}. While there are numerous strategies to address class imbalance in datasets, such as data transformation or training Generative Adversarial Networks (GAN) to generate synthetic data, each approach has its own limitations \cite{zunair2020melanoma, rahman20213c, datta2021soft}. Data transformation, for example, is restricted by the type of dataset, and for MRI images, techniques such as rotation, angle variation, exposure adjustment, or zooming may not be feasible. GANs, on the other hand, typically require substantial amounts of training data, which are often not available, especially to resolve class imbalance, where the minority class may only have a handful of samples \cite{borji2019pros}. Recently, Probabilistic Diffusion Models have gained popularity, particularly with the rise of generative models such as Stable Diffusion \cite{muller2023multimodal}. Unlike traditional methods, diffusion models do not face the challenges associated with data transformation, and many diffusion model architectures perform quite well in generating clinical images, such as MRI or CT scans, even when trained on small datasets \cite{khader2023denoising}.

Most publicly available MRI image datasets are stored in NIfTI format, which requires additional preprocessing software or methods to convert 3D volume data into 2D images \cite{gatidis2022whole, maumet2016sharing}. As a result, researchers often rely on preprocessed image data, typically from only a single view, such as the axial view, while excluding other views like coronal and sagittal. Consequently, the efficacy of deep learning models may be questioned, as these models often capture only a limited portion of the brain to diagnose AD. Furthermore, software tools such as FSL, FreeSurfer, ANTs, and ANTsX require domain-specific knowledge and generally have a steep learning curve, which makes researchers from other domains reluctant to use these tools exclusively for data preprocessing \cite{tustison2021antsx, tustison2024antsx}. Therefore, it is essential to develop a mechanism that simplifies the process of converting 3D volume data to 2D images. 

\begin{figure*}[t!]
    \centering
    \begin{subfigure}[t]{0.275\textwidth}
        \includegraphics[width=\textwidth]{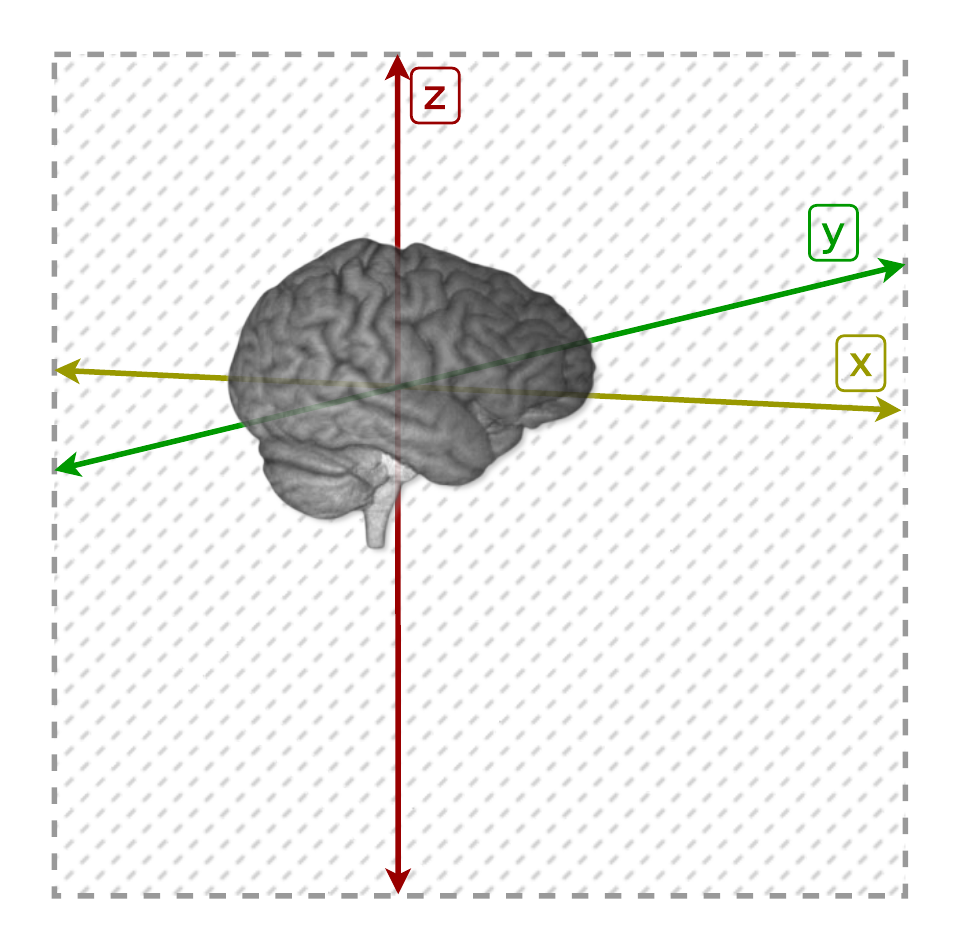}
        \caption{}
        \label{fig:3d-volume}
    \end{subfigure}
    \begin{subfigure}[t]{0.275\textwidth}
        \includegraphics[width=\textwidth]{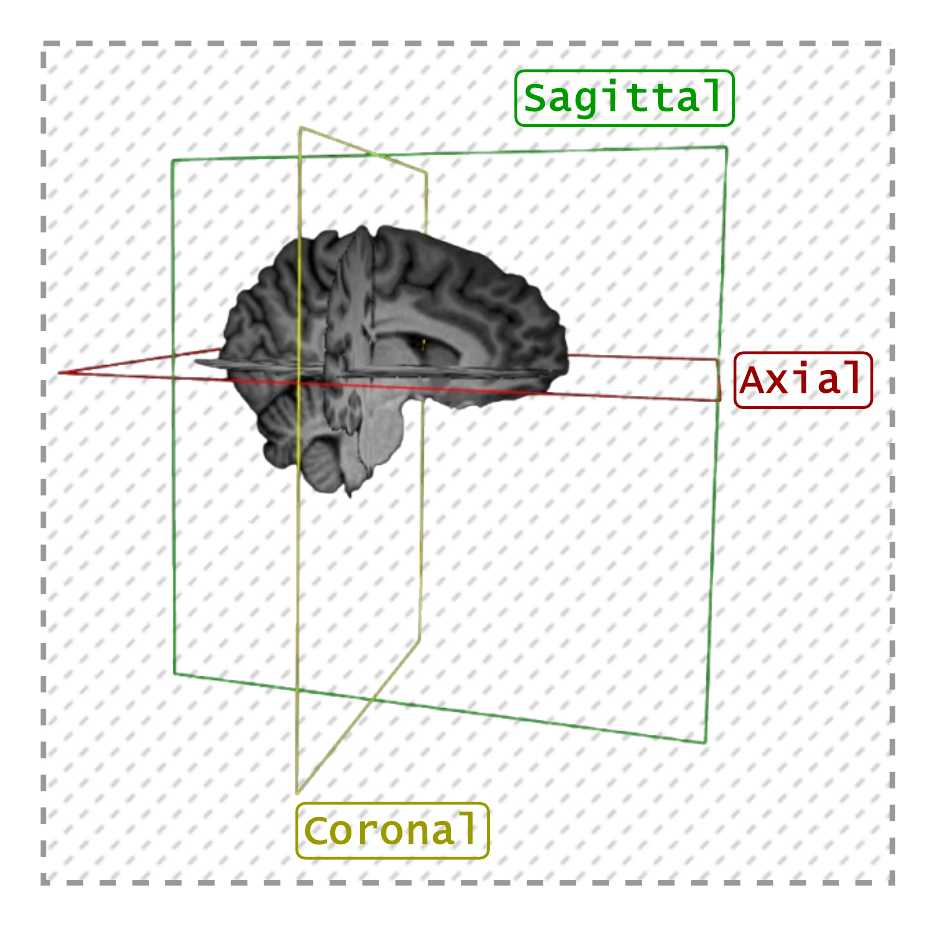}
        \caption{}
        \label{fig:2d-slice}
    \end{subfigure}
    \begin{subfigure}[t]{0.44\textwidth}
        \includegraphics[width=\textwidth]{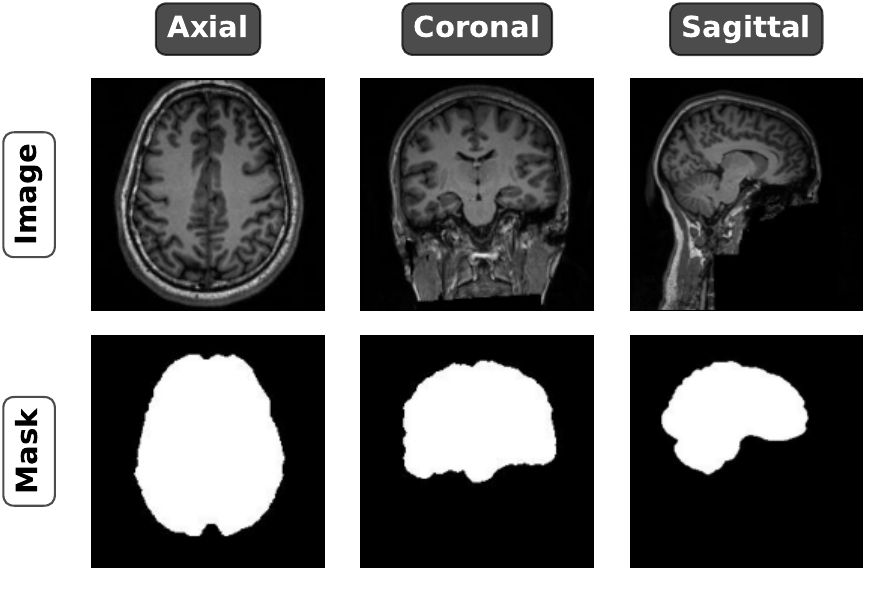}
        \caption{}
        \label{fig:nfbs-dataset}
    \end{subfigure}
    \caption{Subfigure (a) illustrates the 3D MRI volume in a three-dimensional coordinate system, where the brain is represented as a collection of voxels, forming the full anatomical structure. Subfigure (b) shows example 2D slices extracted from the 3D volume, one from each of the three primary anatomical planes: axial (horizontal cross-section), coronal (vertical front-to-back cross-section), and sagittal (side view). Subfigure (c) presents MRI images of the axial, coronal, and sagittal views, along with corresponding brain masks that isolate the brain tissue from surrounding structures.}
    \label{fig:3d-to-2d}
\end{figure*}

Using CNNs to detect AD from MRI images is already prevalent in biomedical research. Yagis \textit{et al.} (2020) proposed 3D CNNs for the diagnosis of AD using structural MRI images, while Cheng \textit{et al.} (2017) introduced multi-domain transfer learning for the early diagnosis of AD \cite{yagis20203d, cheng2017multi}. Guan \textit{et al.} (2021) proposed a multi-instance distillation scheme that transfers knowledge from multi-modal data to an MRI-based network, improving the prediction of mild cognitive impairment conversion, even in data-limited clinical settings \cite{guan2021mri}. Although the methods proposed by these researchers perform well in certain cases of AD detection, the challenge of accurately classifying closely related classes from MRI scans remains a persistent issue for many CNN-based architectures.
 
Recently, quantum machine learning (QML) has emerged as a promising field that combines properties of quantum physics, quantum computing (QC), and classical machine learning (CML) \cite{maheshwari2022quantum}. In CML, training is done on classical computers which depend on bit voltage or charge and relate only to two values: 0 and 1. Logic gates such as AND, OR, and NOT are used to perform operations on these binary values. It is based on classical physics and runs on Boolean algebra. On the other hand, QML is trained on quantum computers, which use quantum bits (qubits) that rely on quantum properties such as the spin of electrons \cite{foletti2009universal}. It can represent not only classical binary states (0 and 1), but also more complex data, including superposition states, where a qubit can simultaneously represent multiple possibilities, and even negative values \cite{jones2001nmr}. This ability to handle a broader range of data and use quantum properties allows quantum computers to perform parallel processing and solve problems with greater efficiency, positioning QML at the forefront of research across various domains, including medical image analysis \cite{qi2023quantum, wei2023quantum}. However, QML models, particularly classical-quantum convolutional neural networks similar to the architecture proposed by Hasan \textit{et al.} \cite{hasan2023bridging}, are still in their infancy. Moreover, the limited access to physical quantum computers and reliance on simulations on classical computers make conducting research in this area challenging. Despite these obstacles, the potential to harness quantum advantage and identify possible challenges associated with it remains a highly active area of research.

In summary, AD represents a major global health crisis, and early diagnosis is essential to slow its progression. Current diagnostic methods, such as PET scans and CSF analysis, are invasive, limiting their accessibility and widespread use. While non-invasive MRI scans can serve as an alternative to PET and CSF analysis for detecting AD, the reliance on preprocessed 2D images to train CNN classifiers and the presence of class imbalance in MRI datasets reduce the efficacy of CML models. This paper addresses these challenges by replacing CML with QML. We then develop a framework to convert 3D MRI data into 2D images. Next, we train a probabilistic diffusion model to generate synthetic images for minority classes and a segmentation model to extract brain tissue relevant to AD detection. Our quantum models are then trained on these processed images. Furthermore, we create a dataset by combining image samples from three anatomical planes to form a 3D representation of the MRI data. 

The key contributions of our work are summarized as follows:

\begin{itemize}
    \item We introduce SkullNet, a multi-view segmentation model trained on the NFBS dataset, which extracts brain tissue and removes the skull and surrounding areas from MRI images.
    \item We train three probabilistic diffusion models, one for the axial plane, one for the coronal plane, and one for the sagittal plane, using samples from the minority class (moderate dementia) of the OASIS-2 dataset. These models are capable of generating synthetic 2D MRI images.
    \item We present, to the best of our knowledge, one of the first end-to-end hybrid classical-quantum convolutional neural network (CQ-CNN) architectures specifically designed for AD detection.
    \item We train classification models based on the CQ-CNN architecture, and our experiments indicate that the \(\beta_8\)-3-qubit model exhibits signs of quantum advantage. It achieves state-of-the-art (SOTA) performance with an accuracy of 97.50\%, using 99.99\% fewer parameters, while our other properly converged models reach similar accuracy in fewer epochs compared to classical models.
\end{itemize}
The remaining sections of this paper are structured as follows. Section \ref{section:2} provides a detailed explanation of the 3D-to-2D data conversion process, along with an overview of the architecture for the proposed diffusion, segmentation, and classification models. Section \ref{section:3} presents the datasets used in this research and outlines their preprocessing steps. In addition, the model training configuration is discussed, and the training progress of the segmentation and generative models is documented. Section \ref{section:4} presents the results of the trained classifiers, evaluates the impact of segmentation, and examines the anomalies encountered during the training of the QML models, along with potential explanations for these issues. This section also includes a comparative analysis with classical SOTA models. Section \ref{section:5} discusses the key findings of the study. Finally, Section \ref{section:6} concludes the paper by summarizing the contributions of this work.

\begin{figure*}[t!]
    \centering
    \includegraphics[width=1.00\textwidth]{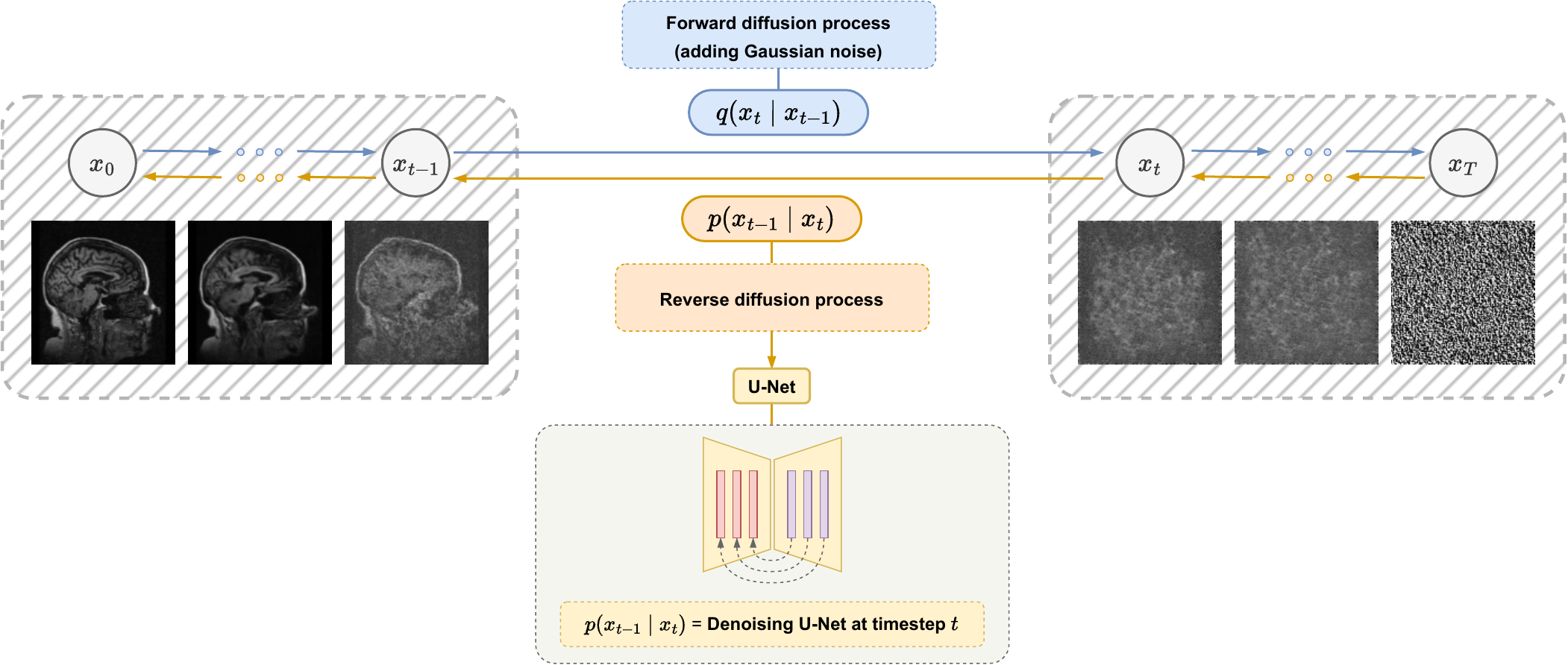}
    \caption{The illustration depicts the diffusion process applied to a sagittal plane of an MRI image sample. The forward process starts with a clean image \( x_0 \) and progressively adds Gaussian noise over \( T \) timesteps. At each timestep, the image is modified according to the conditional distribution \( q(x_t | x_{t-1}) \). As \( t \) increases, the image becomes progressively more corrupted, ultimately resulting in pure noise at \( x_T \). In the reverse process, the model learns to recover the original clean image \( x_0 \) starting from pure noise \( x_T \) by learning the conditional distribution \( p(x_{t-1} | x_t) \) at each timestep using a U-Net. }
    \label{fig:diffusion-arch}
\end{figure*}

\section{Method}
\label{section:2}
\subsection{3D to 2D slice conversion}
Raw MRI data is essentially a 3D volumetric representation, where each voxel corresponds to a small unit of tissue within the scanned area. This 3D data is typically viewed using specialized neuroimaging software, which allows for the exploration of the brain's anatomy from different angles and perspectives. The software helps render the 3D volume, enabling detailed inspection of specific regions, such as gray matter, white matter, or abnormal areas like tumors or lesions. To make this 3D data usable for machine learning purposes, it must first be converted into 2D slices.

To convert 3D MRI data into 2D slices, consider the 3D volume \( V \in \mathbb{R}^3 \), where each point represents a voxel in the scanned region. The data can be visualized from three primary anatomical views: the axial plane (where the \( \mathbb{R}^{xy} \) plane moves along the z-axis), the coronal plane (where the \( \mathbb{R}^{yz} \) plane moves along the x-axis), and the sagittal plane (where the \( \mathbb{R}^{zx} \) plane moves along the y-axis), as shown in Figure \ref{fig:3d-volume} and \ref{fig:2d-slice}. Let \( n \) represent the number of slices to be extracted from each anatomical view, and let \( m \) denote the total number of slices available in that view. The interval between consecutive slices is denoted by \( i \), which determines the spacing between each slice. To calculate the necessary interval \( i \) for extracting \( n \) slices from \( m \) total slices, the following equation is used:
\begin{equation}
   i = \left\lfloor \frac{m}{n} \right\rfloor 
   \label{eq:01}
\end{equation}
where \( \left\lfloor \cdot \right\rfloor \) denotes the floor function, which rounds the slice intervals down to the nearest integer. However, in MRI data, the first and last few slices often do not contain meaningful voxels due to the absence of tissue. Therefore, these slices are excluded after determining the interval \( i \). The total number of valid slices is reduced by \( k_1 \) slices from the beginning and \( k_2 \) slices from the end. The final number of slices to be selected is then:
\begin{equation}
   n_{\text{slices}} = \left\lceil \frac{m}{i} \right\rceil - (k_1 + k_2)
   \label{eq:02-num_of_slices}
\end{equation}
where \( \left\lceil \cdot \right\rceil \) denotes the ceiling function, which rounds the slice number up to the nearest integer. We use Equation \ref{eq:02-num_of_slices} to ensure that the final set of slices extracted from the volume \( V \) is evenly distributed along the anatomical planes \( \mathbb{R}^{xy} \), \( \mathbb{R}^{yz} \), or \( \mathbb{R}^{zx} \), preserving the important structural information from the 3D volume while eliminating irrelevant regions at the edges. We also use this equation as the foundation of our 3D-to-2D data transformation framework, which is written in Python.

In the next step, we move on to the diffusion model architecture, which we use to address class imbalance and oversample images from the minority class.

\begin{figure*}[t!]
    \centering
    \includegraphics[width=1.00\textwidth]{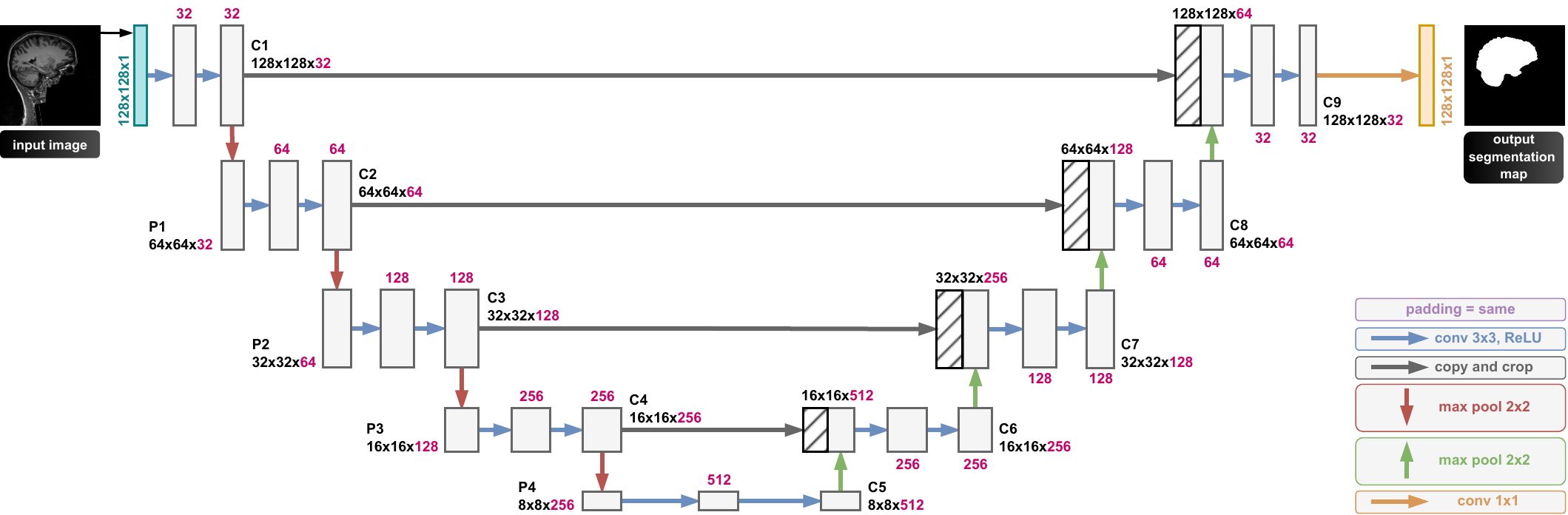}
    \caption{The illustration depicts the architecture of a U-Net, which has four main components: the encoder, bottleneck, decoder, and skip connections. The encoder consists of five convolutional blocks (C1 to C5), progressively reducing spatial dimensions from 128×128 to 8×8 while increasing the feature channels from 32 to 512. Each block uses convolutional operations (blue) with 3×3 kernels, ReLU activations, and ``same" padding (purple), followed by max-pooling layers (red arrows) for downsampling. The bottleneck operates at the lowest resolution (8×8) with the highest abstraction level (512 channels). The decoder upscales feature maps back to 128×128 using transposed convolutions (green arrows) and convolutional blocks (C6 to C9) while reducing feature channels. Skip connections (gray arrows) link corresponding layers of the encoder and decoder, ensuring that fine-grained spatial details are preserved. Finally, a 1×1 convolution (yellow) generates the output segmentation map (128×128×1).}
    \label{fig:u-net}
\end{figure*}

\subsection{Diffusion architecture}

A diffusion model generates data by progressively adding noise to a clean sample and then learning to reverse this process \cite{ho2020denoising}. The forward diffusion process begins with a clean image \( x_0 \) and gradually adds Gaussian noise over \( T \) steps, resulting in a sequence of increasingly noisy images \( x_1, x_2, \dots, x_T \). Mathematically, each step of the forward process is described by the equation:
\begin{equation}
    x_t = \sqrt{\alpha_t} x_{t-1} + \sqrt{1 - \alpha_t} \epsilon_{t-1}, \quad \epsilon_{t-1} \sim \mathcal{N}(0, I) 
    \label{eq:03}
\end{equation}
where \( \alpha_t \) controls the amount of noise added at each step, which is a decreasing sequence approaching zero as $t$ increases, to make the noise progression clearer, \( \epsilon_{t-1} \) represents Gaussian noise sampled from a normal distribution \( \mathcal{N}(0, I) \), and \( I \) is the identity matrix, indicating independent noise components with unit variance. The forward process can be expressed as the conditional distribution \( q(x_t | x_{t-1}) \), which models how the clean image \( x_{t-1} \) transitions to the noisy image \( x_t \):
\begin{equation}
    q(x_t | x_{t-1}) = \mathcal{N}(x_t; \sqrt{\alpha_t} x_{t-1}, (1 - \alpha_t) I)
    \label{eq:04}
\end{equation}
As \( t \) increases, the image becomes progressively noisier until \( x_T \), which is essentially pure noise.

The reverse process aims to recover the clean image \( x_0 \) starting from pure noise \( x_T \). To achieve this, the model learns the conditional distribution \( p(x_{t-1} | x_t) \) at each time step, approximated by a Gaussian distribution:
\begin{equation}
    p(x_{t-1} | x_t) \approx \mathcal{N}(x_{t-1}; \mu_\theta(x_t, t), \Sigma_\theta(x_t, t))
    \label{eq:05}
\end{equation}
Here, \( \mu_\theta(x_t, t) \) and \( \Sigma_\theta(x_t, t) \) represent the predicted mean and variance of the distribution, learned by the neural network during training. In the reverse process, a U-Net is used to estimate the distribution \( p(x_{t-1} | x_t) \). At each step, the noisy image \( x_t \) is passed through the U-Net, which generates a denoised estimate \( \mu_\theta(x_t, t) \) of the previous image \( x_{t-1} \). The U-Net receives both the noisy image \( x_t \) and the timestep \( t \) as inputs, allowing it to progressively remove noise and reconstruct the clean image \( x_0 \). This whole process is illustrated in Figure \ref{fig:diffusion-arch}.

Once synthetic images of the minority class are generated using the diffusion model, we proceed to describe the architecture of the U-Net model, which is not only part of the reverse process of our diffusion model but also used for segmenting the brain tissue region from the MRI images. In the following section, we specifically focus on the U-Net architecture used for our segmentation task.

\subsection{U-Net architecture}
U-Net was originally developed for biomedical image segmentation but has since been widely applied across various fields \cite{ronneberger2015u}. It consists of four main components: encoder, bottleneck, decoder, and skip connections.
\subsubsection*{Encoder}
The encoder is designed to progressively reduce the spatial dimensions of the input image while increasing the number of feature channels. In our architecture, the input to the network is a 128×128 grayscale image, which is processed through a series of convolutional layers. Each convolutional block in the encoder consists of two 3×3 convolution operations with ``same" padding, followed by a ReLU activation function to introduce non-linearity. Downsampling is achieved using max-pooling layers with a 2×2 kernel size, which reduces the spatial dimensions while doubling the number of feature channels. The encoder consists of five convolutional blocks (C1 to C5 in Figure \ref{fig:u-net}), where the number of filters progressively increases from 32 to 512, and the spatial size decreases from 128×128 to 8×8. This hierarchical structure ensures that the network learns abstract and discriminative features while effectively compressing the input image.
\subsubsection*{Bottleneck}
The bottleneck serves as the critical transition point between the encoder and decoder. At this stage, the feature maps achieve the highest level of abstraction while the spatial resolution is at its minimum. In our case, the bottleneck consists of feature maps with a resolution of 8×8 and 512 feature channels, representing a dense and compact abstraction of the input image. This layer captures semantic features that are essential for accurate segmentation while retaining sufficient information for reconstruction during the decoding phase.

\begin{figure*}[t!]
    \centering
    \begin{subfigure}[t]{0.40\textwidth}
        \includegraphics[width=\textwidth]{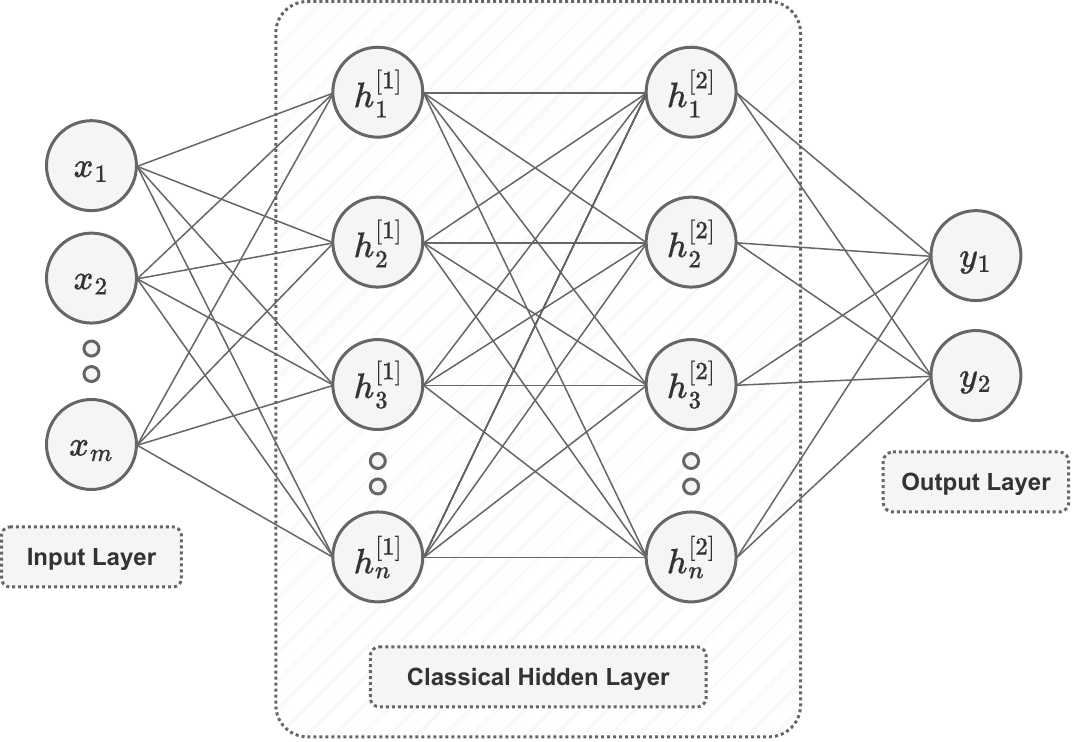}
        \caption{}
        \label{fig:neural-net}
    \end{subfigure}
    \begin{subfigure}[t]{0.59\textwidth}
        \includegraphics[width=\textwidth]{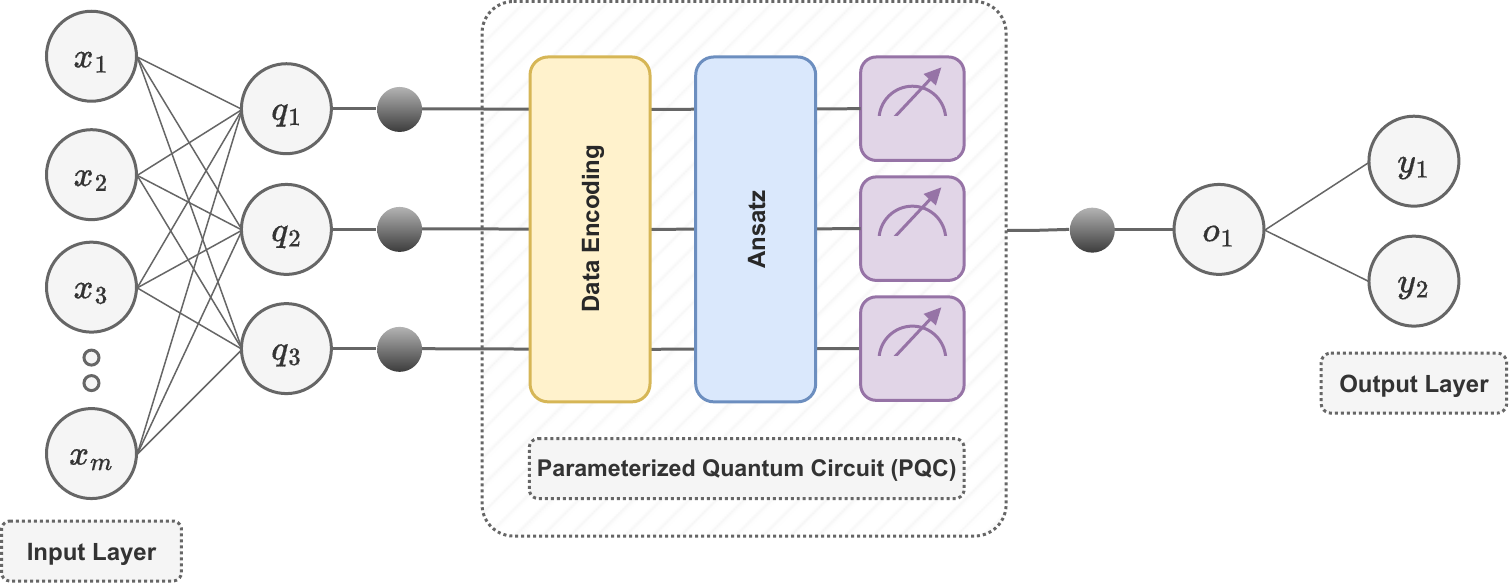}
        \caption{}
        \label{fig:quantum-neural-net}
    \end{subfigure}
    \caption{Schematic depiction of a classical neural network (a) and a quantum neural network (b) for binary classification. In Subfigure (a), \( x_1, x_2, \dots, x_m \) denote the \( m \) input neurons representing the input features. The hidden layer consists of \( n \) neurons represented as \( h_{1}^{[1]}, h_{2}^{[1]}, \dots, h_{n}^{[1]} \), where the superscript \([1]\) indicates the first hidden layer, and the subscript identifies the specific neuron within that layer (e.g., \( h_{1}^{[1]} \) is the first neuron in the first hidden layer). The output layer neurons, representing the predicted probabilities for each class given the input features, are denoted by \( y_1 \) and \( y_2 \). In Subfigure (b), the input and output layers are similar to those in Subfigure (a). However, the classical hidden layers are replaced by a 3-qubit PQC. The classical features are first reduced to match the number of qubits, represented as \( q_1, q_2, q_3 \), with three black dots indicating the qubits. These features are then encoded into quantum states through data encoding. A parameterized ansatz is applied to capture complex relationships using quantum operations. Afterward, quantum measurements are performed, and the PQC outputs a classical probability. This probability passes through an intermediate linear layer, denoted as \( o_1 \). Finally, \( o_1 \) is mapped to the output probability using Equation \ref{eq:12}.}
    \label{fig:nn-qnn}
\end{figure*}

\subsubsection*{Decoder}
The decoder is responsible for restoring the original spatial resolution of the input image while preserving the high-level semantic features extracted by the encoder. The decoder utilizes transposed convolutions (as depicted by the upward green arrows in Figure \ref{fig:u-net}) to upsample the feature maps. Each upsampling step is followed by concatenation with the corresponding encoder feature maps via skip connections, ensuring the retention of fine-grained spatial information. Convolutional layers further refine the upsampled feature maps, progressively reducing the number of feature channels while restoring the spatial dimensions to the original size. For instance, at the final decoding stage, the spatial dimensions are restored to 128×128, with the number of feature channels reduced to 32 (top right of Figure \ref{fig:u-net}).

\subsubsection*{Skip connections}
Skip connections (represented by the gray arrows in Figure \ref{fig:u-net}) are one of the most important features of U-Net. These connections link feature maps from the encoder to the decoder at corresponding resolutions, ensuring that low-level spatial details lost during downsampling are reintroduced into the decoding process. This mechanism helps the network produce sharp and precise segmentation boundaries by combining low-level spatial details with high-level semantic information. Finally, a 1×1 convolutional layer is applied to map the refined feature maps to the desired number of output channels. For binary segmentation tasks, the output is a single-channel 128×128 segmentation map, where each pixel value represents the likelihood of belonging to the target class.

The U-Net architecture we propose here is named SkullNet, as it is specifically designed to segment brain tissues by effectively removing the skull and surrounding structures in MRI images. We use SkullNet to generate variations of datasets so that we can compare and contrast segmented and non-segmented images, both of which are trained using classical-quantum neural network models. In the following section, we provide a detailed description of both approaches.

\begin{figure*}[t!]
    \centering
    \begin{subfigure}[t]{1.0\textwidth}
        \includegraphics[width=\textwidth]{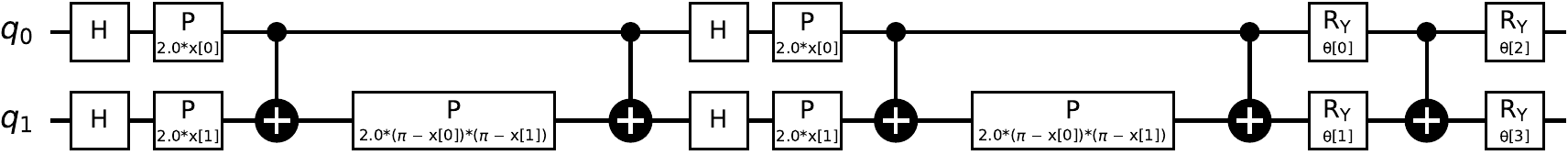}
        \caption{}
        \label{fig:2qubits-qc}
    \end{subfigure}
    \begin{subfigure}[t]{1.0\textwidth}
        \includegraphics[width=\textwidth]{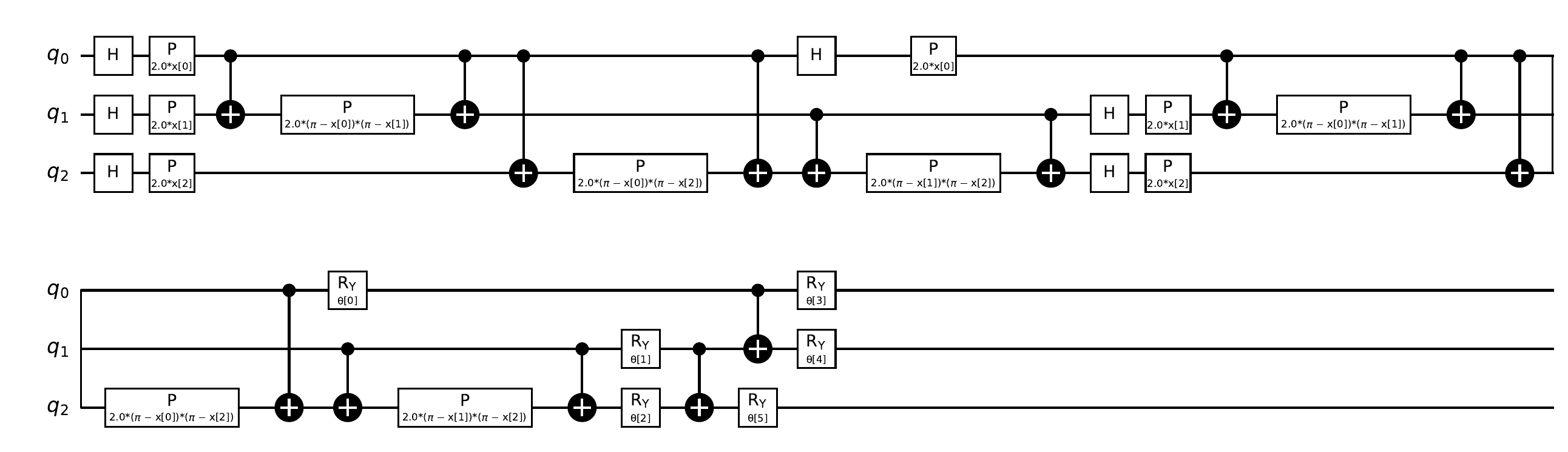}
        \caption{}
        \label{fig:3qubits-qc}
    \end{subfigure}
    \caption{The schematic depicts a PQC using ZZFeatureMap encoding, with subfigure (a) showing a 2-qubit circuit and subfigure (b) showing a 3-qubit circuit. Each qubit is initialized with a Hadamard gate \( H \), followed by phase rotations \( P(2 \cdot x[i]) \) to encode classical data into a quantum state. Entanglement is then introduced through controlled-Z (CZ) gates, which create correlations between qubits by applying phase shifts based on their classical values. A phase rotation \( P(2.0(\pi - x[i])(\pi - x[i])) \) is applied to introduce further phase shifts based on the classical values. The ansatz circuit applies trainable single-qubit rotations \( R_y(\theta_i) \) to further refine the quantum state.}
    \label{fig:qc-circuit}
\end{figure*}

\subsection{Neural network}

\subsubsection{Classical neural network}
A classical neural network is a computational model inspired by the structure of the human brain. It consists of interconnected layers of neurons (or perceptrons), where each neuron processes input data using an activation function (such as ReLU, sigmoid, or tanh) and then transmits the output to the next layer. A classical neural network typically has three types of layers: the input layer, one or more hidden layers, and the output layer (as shown in Figure \ref{fig:neural-net}). 

\subsubsection{Quantum neural network}
A quantum neural network (QNN) is a hybrid classical-quantum machine learning model that integrates quantum mechanics with neural networks. QNNs typically consist of four main components: data encoding, ansatz, quantum measurement, and parameter optimization, as shown in Figure \ref{fig:quantum-neural-net}. The first three components are quantum operations, handled by a quantum computer, while the classical computer optimizes the parameters based on the results from quantum measurements.

\subsubsection*{Data encoding} 
The first step in a QNN is data encoding. Several common encoding techniques, such as angle encoding, amplitude encoding, and basis encoding, are used to map classical data to quantum states. Angle encoding represents classical data as parameters for rotation gates (such as \( R_x \) and \( R_y \)), where the input data directly determine the angles of these gates. Amplitude encoding maps classical data to the amplitudes of quantum states, where the data is represented as a superposition of basis states with complex amplitudes. Basis encoding, on the other hand, assigns classical data directly to specific quantum basis states (such as \( |0\rangle \), \( |1\rangle \), etc.), where each classical value corresponds to a particular state in the computational basis.

ZZFeatureMap is a relatively new encoding technique that extends traditional angle encoding by introducing entanglement between qubits, and is used in our QNN. The process begins with state preparation, where each qubit is initialized in a Hadamard (\(H\)) state, creating a superposition of \( \left| 0 \right\rangle \) and \( \left| 1 \right\rangle \). The feature map then applies parameterized gates \( P(2 \cdot x[i]) \) to each qubit, where \( x[i] \) represents the classical data (as shown in the initial phase of the parameterized quantum circuit (PQC) in Figures \ref{fig:2qubits-qc} and \ref{fig:3qubits-qc}). These gates adjust the phase of each qubit based on the corresponding classical input values.

Next, entanglement is introduced through controlled-Z (CZ) gates, which create correlations between pairs of qubits. This entanglement spreads the classical data across multiple qubits, allowing the quantum system to represent complex correlations that are challenging for classical models to capture. Mathematically, the encoding process using the ZZFeatureMap for a \(N\)-qubit system can be expressed as:
\begin{equation}
    \left| \psi(\mathbf{x}) \right\rangle = H^{\otimes N} \cdot \prod_{i=1}^N P(2 \cdot x[i]) \cdot \prod_{i<j} \text{CZ}(i,j)
    \label{eq:06}
\end{equation}
where \( H^{\otimes N} \) represents the Hadamard operation applied to each qubit, \( P(2 \cdot x[i]) \) is the parameterized gate acting on each qubit, and \( \text{CZ}(i,j) \) is the controlled-Z gate applied between qubits \( i \) and \( j \), creating entanglement. This results in an entangled state \( \left| \psi(\mathbf{x}) \right\rangle \), which encodes the classical data into a quantum state.

\begin{figure*}[t!]
    \centering
    \includegraphics[width=1.00\textwidth]{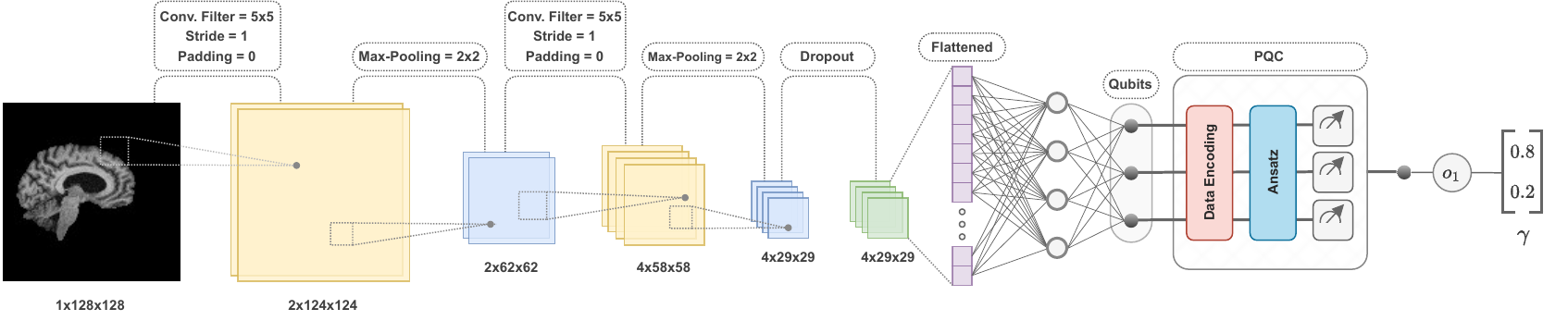}
    \caption{The illustration depicts a hybrid classical-quantum neural network architecture for binary image classification. The input is a grayscale 2D MRI slice of size 1x128x128, which passes through a convolutional layer with a 5x5 filter, a stride of 1, and no padding, producing 2x124x124 feature maps, followed by 2x2 max-pooling, which reduces it to 2x62x62. A second convolutional layer with the same filter settings generates 4x58x58 feature maps, which are then reduced to 4x29x29 through max-pooling. A dropout layer is applied for regularization, and the output is flattened for the fully connected (dense) layer. The processed data is then fed into a PQC, where classical data is encoded into quantum states, followed by ansatz layers with learnable parameters updated using the gradient descent algorithm defined in Equation \ref{eq:11}, and finally measured to produce classification probabilities, resulting in the output vector \(\gamma\).}
    \label{fig:qcnn-arch}
\end{figure*}

\subsubsection*{Ansatz}
Following data encoding, the output quantum state \( \left| \psi(\mathbf{x}) \right\rangle \) is passed as input to the ansatz. The ansatz applies a sequence of trainable quantum gates to change the encoded quantum state, allowing it to learn patterns for making predictions. In ZZFeatureMap encoding, the ansatz uses parameterized rotation gates \( R_y(\theta_i) \) (depicted at the end of the PQC in Figures \ref{fig:2qubits-qc} and \ref{fig:3qubits-qc}), where \( \theta_i \) represents a trainable parameter for the \( i \)-th qubit. For an \( N \)-qubit system, the ansatz is formulated as:  
\begin{equation}
    \left| \psi(\theta) \right\rangle = U(\theta) \cdot \left| \psi(x) \right\rangle = \prod_{i=1}^{N} R_y(\theta_i) \cdot \left| \psi(x) \right\rangle   
    \label{eq:07}
\end{equation}
where \( U(\theta) \) represents the parameterized ansatz circuit, and the product notation indicates the sequential application of rotation gates to all \( N \) qubits. 

\subsubsection*{Quantum measurement}
Once the ansatz circuit has transformed the quantum state, the next step is quantum measurement. Measurement collapses the quantum state to one of the eigenstates of the measurement operator, which in the case of QNNs is the Pauli-Z operator \( \sigma_z \), representing a computational basis measurement. The measurement results give classical probabilities that can be used to compute the output of the quantum neural network. The probability \( p_i \) of obtaining a specific measurement outcome \( i \) is given by:

\begin{equation}
    p_i = \left| \langle i | \psi(\theta) \rangle \right|^2
    \label{eq:08}
\end{equation}

where \( |i\rangle \) represents the \( i \)-th eigenstate, and \( \left| \psi(\theta) \right\rangle \) is the quantum state after the ansatz transformation. The expected value of the measurement outcome \( M \) can be computed as:

\begin{equation}
    \langle M \rangle = \sum_i l_i \cdot p_i = \sum_i l_i \left| \langle i | \psi(\theta) \rangle \right|^2
    \label{eq:09}
\end{equation}

where \( l_i \) is the eigenvalue associated with the eigenstate \( |i\rangle \), typically \( +1 \) or \( -1 \) for Pauli-Z measurements.

\subsubsection*{Parameter optimization}
The final step in the QNN process is parameter optimization. The parameters \( \theta = (\theta_0, \theta_1, \theta_2, \dots, \theta_m) \) of the ansatz circuit are optimized classically to minimize a loss function \( L(\theta) \), which is based on the measured outcomes of the quantum circuit. The loss function used in our QNN for classification tasks is the cross-entropy loss:
\begin{equation}
    L(\theta) = - \frac{1}{N} \sum_{j=1}^N \sum_{c=1}^C y_{jc} \log(p_i = c)   
    \label{eq:10}
\end{equation}
where \( N \) is the number of samples, \( C \) is the number of classes, \( y_{jc} \) is the true label, and \( p_i = c \) is the probability of measuring the eigenstate corresponding to class \( c \). 

The classical optimization algorithm, known as gradient descent, is used to update the parameters of the ansatz circuit:
\begin{equation}
    \theta_i^{(k+1)} = \theta_i^{(k)} - \eta \cdot \frac{\partial L(\theta)}{\partial \theta_i}
    \label{eq:11}
\end{equation}
where \( \eta \) is the learning rate, and the gradient \( \frac{\partial L(\theta)}{\partial \theta_i} \) is computed using the parameter-shift rule on the quantum device. 

The output of the PQC, \(o_1\), is a classical probability value, which is then mapped to the output probability, \(\gamma\), using the following equation:  
\begin{equation}
    \gamma = \text{concatenation}((o_1, 1 - o_1), -1)
    \label{eq:12}
\end{equation}
where the concatenation operation combines \(o_1\) with \(1 - o_1\) to form the final output vector \(\gamma\)  (as shown in the output of Figure \ref{fig:qcnn-arch}).

\subsubsection{Convolutional neural network}
In tasks like image classification, such as detecting AD from MRI images, convolutional neural networks (CNNs) are commonly used. Unlike traditional neural networks, CNNs use specialized layers called convolutional filters to process input data and detect local features such as edges, textures, and shapes. These features are then passed through activation functions and processed by pooling layers, which reduce the spatial dimensions of the feature maps while retaining the most important information. The features are then flattened into a one-dimensional vector and fed into a fully connected layer, which generates the output. For classification tasks, this output is usually passed through a softmax function, which converts the raw output into a probabilistic distribution, where each class is assigned a probability between 0 and 1, and the sum of all probabilities equals 1. 

A classical CNN can be transformed into a hybrid classical-quantum convolutional neural network (CQ-CNN) by incorporating a PQC after the flattened one-dimensional vector.  To ensure compatibility, we reduce the number of neurons in the fully connected layer so that its connections match the number of qubits in the PQC. In our CQ-CNN architecture, we also replace the softmax layer with Equation \ref{eq:12} to generate the final output probabilities. In the CQ-CNN, as illustrated in Figure \ref{fig:qcnn-arch}, the convolutional filters first extract local features from the input MRI slice, which are then processed through ReLU activation functions and max-pooling layers. The resulting feature maps are flattened into a one-dimensional vector and passed through a fully connected layer with a reduced number of neurons. The output is then fed into the PQC, where classical data is encoded into quantum states, processed through quantum operations, followed by measurement and classical optimization. The measured output is then passed through a final one-dimensional classical linear layer to produce the classification probability \( \gamma \).

We apply this hybrid CQ-CNN architecture in our experiments to train classifiers on both segmented and non-segmented MRI images for AD detection. To ensure consistency, we maintain a constant number of trainable parameters across all experiments. The CQ-CNN architecture has only 13.7K trainable parameters, which is extremely low compared to modern classical CNN models such as ResNet and DenseNet \cite{yaqoob2021resnet, huang2018condensenet}. We intentionally keep the parameter count low to accurately evaluate the power of PQC, assess the feasibility of achieving quantum advantage, and identify potential challenges associated with it.

\section{Experiments}
\label{section:3}
\subsection{Dataset}
Our experiments involve three key steps: training diffusion models for the minority class in each anatomical plane, developing a segmentation model to extract brain tissues and remove skulls and other irrelevant parts from MRI scans, and finally training classification models using the processed image datasets. For training the segmentation model, we use the NFBS dataset \cite{eskildsen2012beast} (with sample images shown in Figures \ref{fig:3d-volume}, \ref{fig:2d-slice}, \ref{fig:nfbs-dataset}, and \ref{fig:u-net}), while the OASIS-2 dataset \cite{marcus2010open} (with processed samples shown in Figures \ref{fig:diffusion-arch} and \ref{fig:qcnn-arch}) is utilized for both the diffusion and classification models. Both datasets consist of T1-weighted MRI scans. However, the OASIS-2 dataset contains exclusively 3D MRI volumes, whereas the NFBS dataset includes 3D volumes along with brain tissue masks and segmented images. This combination makes the NFBS dataset ideal for evaluating our framework (Equation \ref{eq:02-num_of_slices}), which converts 3D volumes into 2D images. 

The OASIS-2 dataset consists of four classes: non-demented, very mild demented, mild demented, and moderate demented. The class distribution is highly imbalanced, with non-demented samples being overrepresented and moderate demented samples underrepresented. To address this issue, we use our diffusion model to generate synthetic data for the minority class. Lastly, since our CQ-CNN architecture is designed for binary classification, we exclude the very mild and mild demented samples and perform our experiments on the non-demented and moderate demented classes.

\begin{table}[ht]
    \centering
    \renewcommand{\arraystretch}{1.0}
    \setlength{\tabcolsep}{10pt} 
    \begin{tabular}{lcccccc}
         \toprule
         Plane & \(m\) & \(n\) & \(i\) & \(k_1\) & \(k_2\) & \(n_{\text{slices}}\) \\
         \midrule
         Axial    & 256 & 40 & 6 & 10 & 18 & \textbf{15} \\
         Coronal  & 256 & 40 & 6 & 10 & 18 & \textbf{15} \\
         Sagittal & 192 & 40 & 4 & 13 & 15 & \textbf{20} \\
         \bottomrule
    \end{tabular}
    \caption{3D-to-2D slice extraction for each anatomical plane in the NFBS dataset. Here, \(m\) represents the total number of slices available in the plane, \(n\) denotes the initial number of slices to be extracted (calculated using Equation \ref{eq:01}), \(i\) is the interval between consecutive slices, \(k_1\) and \(k_2\) are the numbers of slices to be excluded from the beginning and end, respectively, and \(n_{\text{slices}}\) (calculated using Equation \ref{eq:02-num_of_slices}) is the final number of valid slices extracted per sample in each plane.}
    \label{tab:nfbs_slice}
\end{table}

\subsection{Model configuration and training}

\subsubsection*{Segmentation model}
We begin our experiments by training our segmentation model, SkullNet, which is based on a U-Net architecture, as illustrated in Figure \ref{fig:u-net}. The process starts with data preprocessing, where we apply our 3D-to-2D conversion framework to transform the 3D volumetric images from the NFBS dataset into 2D slices. For each 3D volume, we extract 15 slices from both the axial and coronal planes and 20 slices from the sagittal plane. The extraction procedure is detailed in Table \ref{tab:nfbs_slice}.

\begin{figure}[b]
    \centering
    \includegraphics[width=1.0\linewidth]{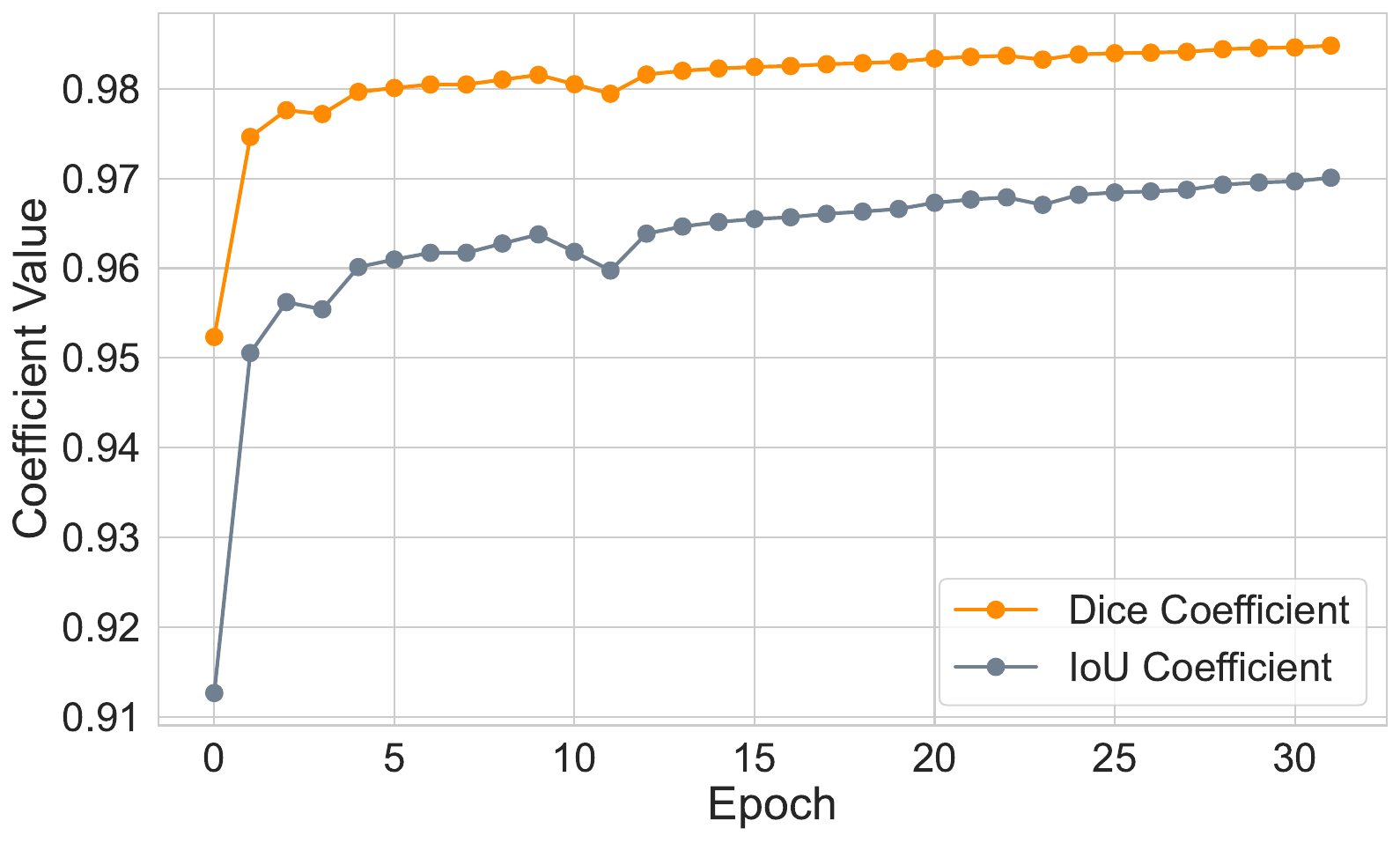}
    \caption{The graph depicts the training progress of the SkullNet model, showing the Dice and IoU coefficients over 30 epochs. The Dice coefficient (orange) increases rapidly and stabilizes around 0.985, while the IoU coefficient (gray) converges to around 0.97.}
    \label{fig:skullnet-train}
\end{figure}

\begin{figure*}[ht]
    \centering
    \begin{subfigure}[t]{1.0\textwidth}
        \includegraphics[width=\textwidth]{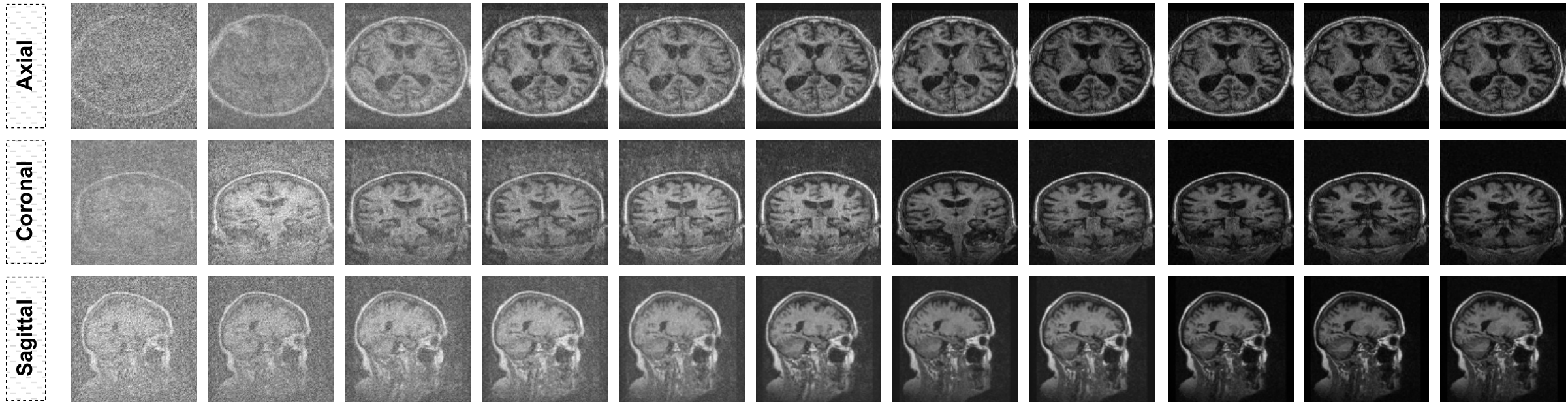}
    \end{subfigure}
    \begin{subfigure}[t]{1.0\textwidth}
        \includegraphics[width=\textwidth]{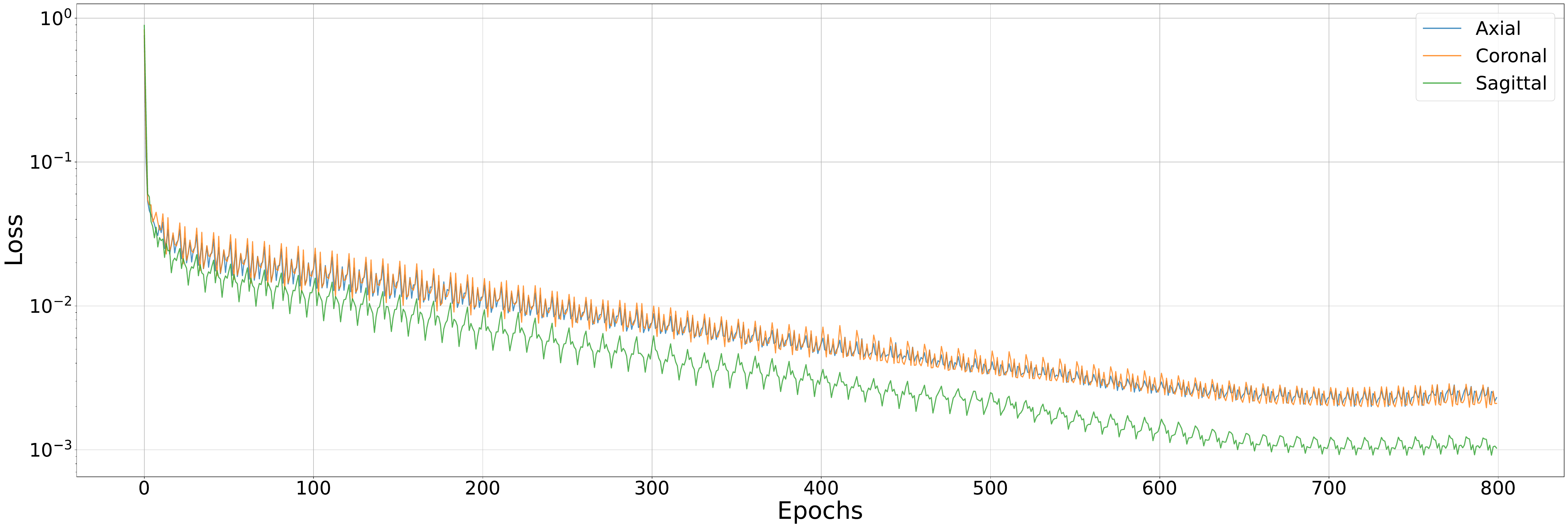}
    \end{subfigure}
    \caption{The visuals present the training loss curves over 800 epochs for three distinct diffusion models, each designed to generate MRI images from the axial, coronal, and sagittal planes. The upper section displays the progression of generated images at different stages of training, showcasing the refinement of details as training advances. The lower graph presents the training loss curves for the three models. The y-axis, shown on a logarithmic scale, highlights the sharp decline in loss during the early stages of training. All three models follow a similar convergence pattern, with losses stabilizing around 700 epochs. Despite reaching an apparent plateau, minor fluctuations persist, indicating that ongoing training continues to improve the models' image generation quality.}
    \label{fig:diffusion-train}
\end{figure*}

The dataset consists of 125 MRI scans for each anatomical plane. Therefore, there are \(125 \times 15 = 1,875\) slices per plane for both the axial and coronal planes, and \(125 \times 20 = 2,500\) slices from the sagittal plane, resulting in a total of 6,250 2D images. To construct the test set, 105 slices are randomly selected from each of the axial and coronal planes, and 140 slices from the sagittal plane, totaling 350 images for the test set. The remaining slices are allocated to the training set, which consists of 1,770 axial and coronal images per plane, and 2,360 sagittal images, totaling 5,900 images in the training set. The corresponding brain masks are also extracted for each sample in both the train and test sets. To ensure consistency, all images are resized to 128×128 pixels before training the SkullNet model, and this resolution is maintained in subsequent experiments. Since the preprocessed NFBS dataset contains images and their segmentation masks for all three anatomical planes, the trained SkullNet model is capable of segmenting brain tissue in axial, coronal, and sagittal planes. However, the orientation of the images must strictly correspond to the views shown in Figure \ref{fig:nfbs-dataset}. 

\begin{table*}[htbp]
    \centering
    \caption{Performance analysis of classification models across axial, coronal, sagittal, and combined 3-plane views. Key evaluation metrics, including precision, F1-score, specificity, accuracy, and training time, are provided for models using both 2-qubit (\(\alpha_i\)) and 3-qubit (\(\beta_i\)) configurations, where \(i\) represents experiments conducted on a specific dataset variation. Each metric is reported as the mean and standard deviation over multiple runs. The analysis also examines the impact of skull-stripping (denoted by \(\Xi\)) on model performance and compares results based on whether the models were trained with single-plane (2D) or multi-plane (3D) images. Boldface numbers indicate the best performance. The symbol \(\uparrow\) denotes that a higher value is better, while \(\downarrow\) signifies that a lower value is better.}
    \renewcommand{\arraystretch}{1.2}
    \resizebox{1.0\textwidth}{!}{
    \begin{tabular}{
    l@{\hspace{10pt}}
    l@{\hspace{10pt}}
    c@{\hspace{10pt}}
    c@{\hspace{10pt}}
    c@{\hspace{10pt}}
    l@{\hspace{20pt}}
    l@{\hspace{20pt}}
    l@{\hspace{20pt}}
    l@{\hspace{10pt}}
    c}
    \hline 
    & & & & & & & & & hh:mm:ss \(\pm\) mm:ss \\
    \cmidrule{10-10}
    & \textbf{Plane} & \textbf{\(\Xi\)} & \textbf{Dim.} & \textbf{Qubits} & \textbf{Precision [\(\uparrow\)]} & \textbf{F1-Score [\(\uparrow\)]} &\textbf{Specificity [\(\uparrow\)]} & \textbf{Accuracy [\(\uparrow\)]} & \textbf{Training Time [\(\downarrow\)]} \\
    \toprule
    \(\ \alpha_1 \) &
    Axial & \(\times\) & 2D & 2 & 0.8777 \(\pm\) 0.0492 & 0.9345 \(\pm\) 0.0279 & 0.9843 \(\pm\) 0.0052 & 0.9840 \(\pm\) 0.0075 & 00:08:20 \(\pm\) 00:17\\
    \(\ \alpha_2 \) & Axial & \(\checkmark\) & 2D & 2 & \textbf{0.8955 \(\pm\) 0.1235} & \textbf{0.9342  \(\pm\) 0.0569} & \textbf{0.9853 \(\pm\) 0.0180} & \textbf{0.9851 \(\pm\) 0.0136} & 00:07:17 \(\pm\) 00:43 \\
    \cmidrule{6-10}
    \(\ \alpha_3 \) & Coronal & \(\times\) & 2D & 2 & \textbf{0.9471 \(\pm\) 0.0307} & \textbf{0.9727 \(\pm\) 0.0162} & \textbf{0.9930 \(\pm\) 0.0043} & \textbf{0.9937 \(\pm\) 0.0038} & 00:04:24 \(\pm\) 00:35 \\
    \(\ \alpha_4 \) & Coronal & \(\checkmark\)  & 2D & 2 & 0.6931 \(\pm\) 0.1474 & \textbf{0.8143 \(\pm\) 0.1032} &  \textbf{0.9405 \(\pm\) 0.0395} & \textbf{0.9471 \(\pm\) 0.0351} & 00:04:14 \(\pm\) 00:50 \\
    \cmidrule{6-10}
    \(\ \alpha_5 \) & Sagittal & \(\times\) & 2D & 2 & \textbf{0.9569 \(\pm\) 0.0610} & \textbf{0.9775 \(\pm\) 0.0318} & \textbf{0.9941 \(\pm\) 0.0083} & \textbf{0.9948 \(\pm\) 0.0074} & 00:04:28 \(\pm\) 00:44\\
    \(\ \alpha_6 \) & Sagittal & \(\checkmark\)  & 2D & 2 & 0.7068 \(\pm\) 0.2709 & 0.8088 \(\pm\) 0.1947 & 0.9404 \(\pm\) 0.0642 & 0.9370 \(\pm\) 0.0713 & 00:04:51 \(\pm\) 00:21\\
    \cmidrule{6-10}
    \(\ \alpha_7 \) & 3-Plane & \(\times\) & 3D & 2 & \textbf{0.9034 \(\pm\) 0.1129} & 0.9246 \(\pm\) 0.0489 & 0.9861 \(\pm\) 0.0169 & 0.9823 \(\pm\) 0.0125 & 00:25:18 \(\pm\) 02:12\\
    \(\ \alpha_8 \) & 3-Plane & \(\checkmark\) & 3D & 2 & 0.6721 \(\pm\) 0.1411 & 0.7527 \(\pm\) 0.2505 & 0.9570 \(\pm\) 0.0136 & 0.9350 \(\pm\) 0.0390 & 00:24:57 \(\pm\) 04:06\\ 
    \midrule
    \(\ \beta_1 \) & Axial & \(\times\) & 2D & 3 & \textbf{0.9069 \(\pm\) 0.0080} & \textbf{0.9512 \(\pm\) 0.0043} & \textbf{0.9872 \(\pm\) 0.0012} & \textbf{0.9886 \(\pm\) 0.0011} & 00:22:40 \(\pm\) 03:10 \\
    \(\ \beta_2 \) & Axial & \(\checkmark\) & 2D & 3 & 0.7883 \(\pm\) 0.0945 & 0.8740 \(\pm\) 0.0677 & 0.9686 \(\pm\) 0.0167 & 0.9701 \(\pm\) 0.0175 & 00:18:43 \(\pm\) 01:13 \\
    \cmidrule{6-10}
    \(\ \beta_3 \) & Coronal & \(\times\) & 2D & 3 & 0.9186 \(\pm\) 0.0096 & 0.9575 \(\pm\) 0.0052 & 0.9895 \(\pm\) 0.0022 & 0.9896 \(\pm\) 0.0005 & 00:10:52 \(\pm\) 01:41 \\
    \(\ \beta_4 \) & Coronal & \(\checkmark\) & 2D & 3 & \textbf{0.7196 \(\pm\) 0.3502} & 0.8123 \(\pm\) 0.2418 & 0.9283 \(\pm\) 0.0955 & 0.9362 \(\pm\) 0.0849 & 00:10:11 \(\pm\) 01:00 \\
    \cmidrule{6-10}
    \(\ \beta_5 \) & Sagittal & \(\times\) & 2D & 3 & 0.9492 \(\pm\) 0.0719 & 0.9094 \(\pm\) 0.0524 & 0.9929 \(\pm\) 0.0100 & 0.9811 \(\pm\) 0.0089 & 00:10:15 \(\pm\) 00:52 \\ 
    \(\ \beta_6 \) & Sagittal & \(\checkmark\) & 2D & 3 & \textbf{0.7814 \(\pm\) 0.2334} & \textbf{0.8676 \(\pm\) 0.1484} & \textbf{0.9575 \(\pm\) 0.0501} & \textbf{0.9622 \(\pm\) 0.0445} & 00:10:07 \(\pm\) 00:20 \\
    \cmidrule{6-10}
    \(\ \beta_7 \) & 3-Plane & \(\times\) & 3D & 3 & 0.9023 \(\pm\) 0.0337 & \textbf{0.9485 \(\pm\) 0.0186} & \textbf{0.9864 \(\pm\) 0.0052} & \textbf{0.9879 \(\pm\) 0.0046} & 01:23:05 \(\pm\) 27:41 \\
    \(\ \beta_8 \) & 3-Plane & \(\checkmark\) & 3D & 3 & \textbf{0.8319 \(\pm\) 0.0686} & \textbf{0.8945 \(\pm\) 0.0257} & \textbf{0.9755 \(\pm\) 0.0126} & \textbf{0.9750 \(\pm\) 0.0076} & 01:20:55 \(\pm\) 24:41 \\
    \bottomrule
    \end{tabular}
    }
    \label{tab:cls-models}
\end{table*}
Training of the SkullNet model, shown in Figure \ref{fig:skullnet-train}, is evaluated using the IoU coefficient, which quantifies the overlap between the predicted and ground-truth masks, and the Dice coefficient, which assesses segmentation accuracy based on precision and recall. \footnote{The segmentation and generative models, along with the relevant experimental code, are available at: \href{https://github.com/mominul-ssv/alz-cq-cnn}{https://github.com/mominul-ssv/alz-cq-cnn}.}

\subsubsection*{Generative model}
The diffusion model is trained using the OASIS-2 dataset, which acts as the primary dataset for our experiments with classification models. The OASIS-2 dataset consists of 3D volumetric MRI scans, similar to those in the NFBS dataset. Therefore, we apply the same 3D-to-2D conversion method to extract 2D slices and subsequently partition the data into a 90:10 train-test split. Since the purpose of the diffusion model is to address class imbalance, we focus on the minority class in the training set, which is the moderate dementia class. All available images from this class are used to train the diffusion model.

Since 2D slices are extracted from the axial, coronal, and sagittal planes, three variations of the processed datasets are created. Separate diffusion models are trained on the images of the minority class for each plane (with the progression of the training shown in the upper section and the corresponding loss curves presented in the lower section of Figure \ref{fig:diffusion-train}). Following training, the class imbalance in the training set is mitigated by generating synthetic images to augment the minority class. However, the test set remains unchanged. Although the test set is still imbalanced, this approach ensures that the classification models are evaluated on authentic, real-world images rather than synthetic data.

In addition, we create three additional variations of the OASIS-2 dataset, where all images in the training and test sets are segmented using the SkullNet model. This experiment aims to evaluate whether our trained models can effectively classify brain tissue relevant to AD. Lastly, we combine all images from the axial, coronal, and sagittal planes to create two final variations of the dataset: one segmented and one non-segmented. These combined datasets effectively form a 3-plane 3D dataset, and models trained on them can classify images from all three anatomical planes.

\subsubsection*{Classification model}
The classification models (with architecture illustrated in Figure \ref{fig:qcnn-arch}) are trained using all variations of the processed OASIS-2 dataset. Since the PQC of our classification model is trained on a quantum computer, we simulate quantum computations on classical computers using Qiskit \cite{QiskitCommunity2017}. We conduct experiments with both 2-qubit and 3-qubit circuits. In addition, we perform experiments on pure classical models with the same number of parameters as the CQ-CNN architecture and compare their training convergence rates with those of the quantum models.

\section{Results}
\label{section:4}
\subsection{Performance analysis}
The performance of classification models with various qubit configurations, trained on both skull-stripped and non-skull-stripped datasets across different MRI planes, is summarized in Table \ref{tab:cls-models}. From this table, we observe the following.

\textbf{Effect of skull-stripping:} Models trained on skull-stripped datasets generally achieve lower scores across evaluation metrics compared to those trained on non-skull-stripped datasets. For example, the \(\alpha_6\) model achieves an F1 score of 0.8088, whereas the \(\alpha_5\) model attains a significantly higher score of 0.9775. However, despite their lower numerical performance, skull-stripped models provide more clinically reliable predictions, as their outputs are derived solely from brain tissue directly relevant to AD.   

\textbf{Effect of qubits:} Unlike classical CNNs, where increasing the number of parameters typically enhances performance, quantum models do not always benefit from additional qubits. While a larger quantum system enables the model to capture more complex patterns, it also increases sensitivity to quantum noise, which can degrade performance. This is evident in the 2-qubit models $\alpha_3$ and $\alpha_5$, which achieve F1-scores of 0.9727 and 0.9775, compared to the 3-qubit models $\beta_3$ and $\beta_5$, with lower scores of 0.9575 and 0.9094. However, an opposite trend is observed in the 3-qubit models $\beta_6$ and $\beta_8$, which achieve F1-scores of 0.8676 and 0.8945, both higher than their 2-qubit counterparts, $\alpha_6$ and $\alpha_8$, which score 0.8088 and 0.7527. This suggests that, in certain cases, 3-qubit models can make use of their additional qubits more effectively to capture patterns in AD-relevant brain tissues compared to 2-qubit models.

\begin{figure*}[t!]
    \centering
    \includegraphics[width=1.00\textwidth]{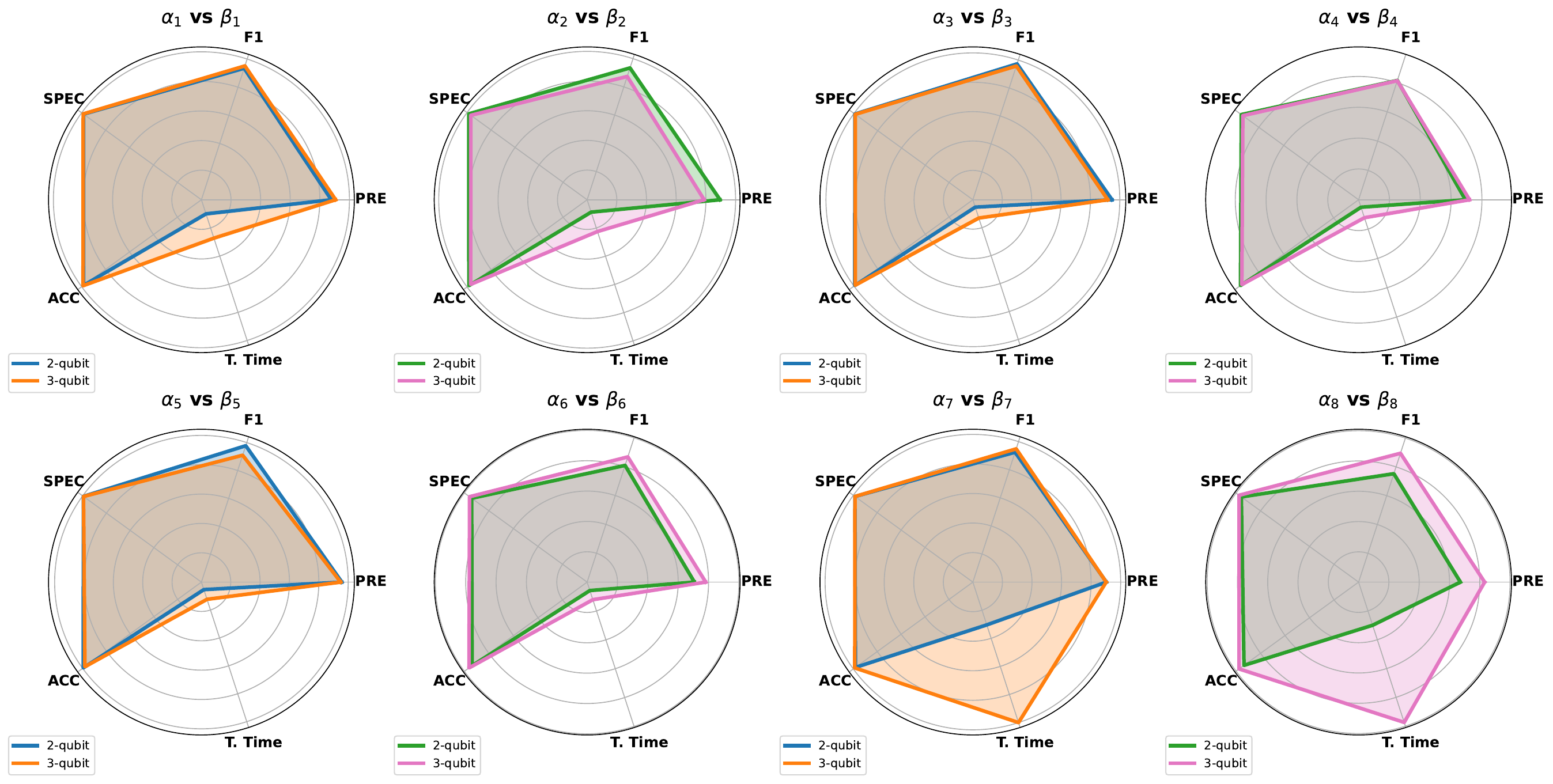}
    \caption{Radar plots compare the performance of models with different qubit configurations across evaluation metrics: accuracy (ACC), specificity (SPEC), F1-score (F1), precision (PRE), and training time (T. Time). Each subplot represents a comparison between the 2-qubit model (\(\alpha_i\)) and its corresponding 3-qubit model (\(\beta_i\)), where both models are trained on the same dataset \(i\). The radar plots highlight that despite the use of 3-qubit models (e.g., \(\alpha_7\) vs. \(\beta_7\)), the overall performance improvements are minimal. In contrast, training time increases significantly with the addition of qubits.}
    \label{fig:radar-plot}
\end{figure*}

\textbf{Trade-off between time and performance:} While increasing the number of qubits may occasionally improve performance, overall gains remain limited. This observation is detailed in the radar plots in Figure \ref{fig:radar-plot}, where the 2-qubit \(\alpha_i\) models and their corresponding 3-qubit \(\beta_i\) models from Table \ref{tab:cls-models} show similar area coverage. The primary difference is the significant increase in training time, as quantum models scale computationally with circuit depth. For example, training the 3-qubit model \(\beta_8\) takes 1 hour and 20 minutes, nearly four times longer than the 24 minutes required for the 2-qubit model \(\alpha_8\). A similar trend is observed across other 3-qubit models, where adding qubits doubles or triples the training time without providing proportional performance improvements.

\begin{figure*}[ht!]
    \centering
    \begin{subfigure}[t]{1.0\textwidth}
        \includegraphics[width=\textwidth]{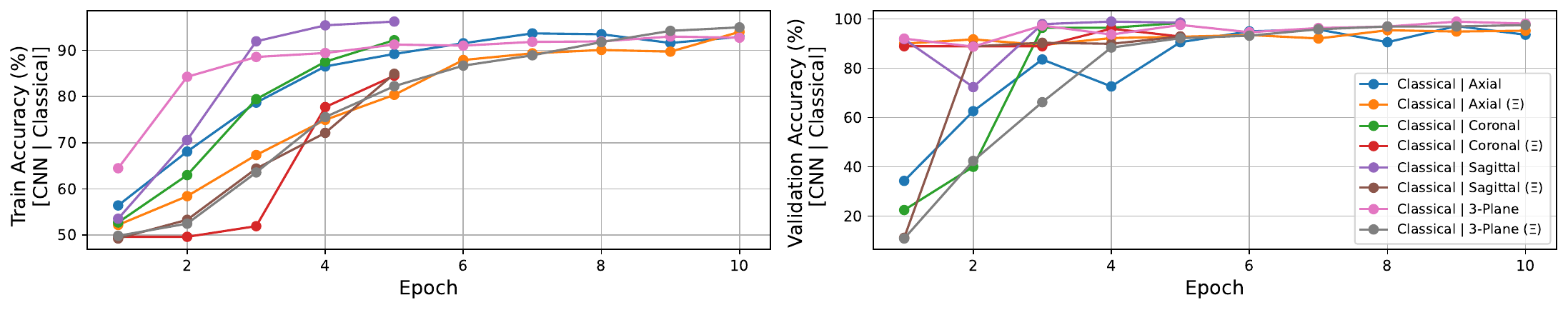}
    \end{subfigure}
    \begin{subfigure}[t]{1.0\textwidth}
        \includegraphics[width=\textwidth]{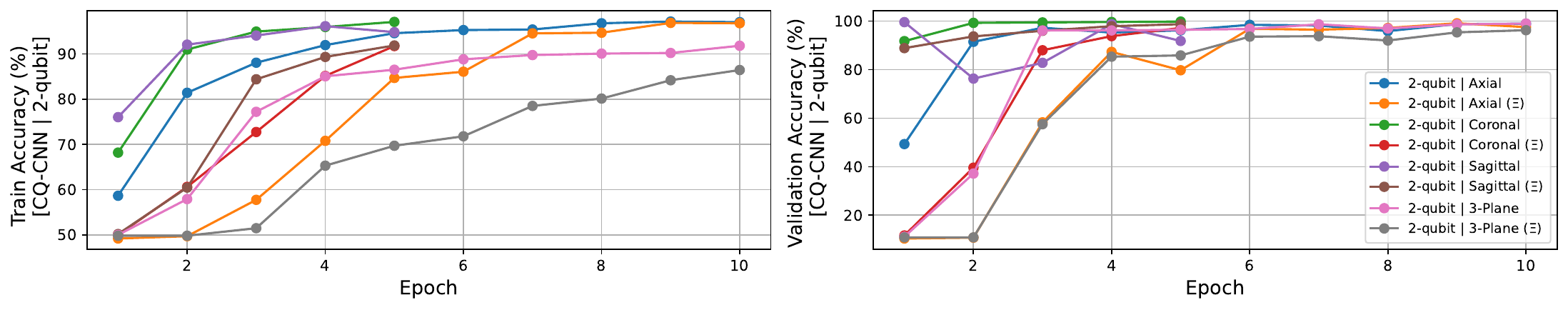}
    \end{subfigure}
    \begin{subfigure}[t]{1.0\textwidth}
        \includegraphics[width=\textwidth]{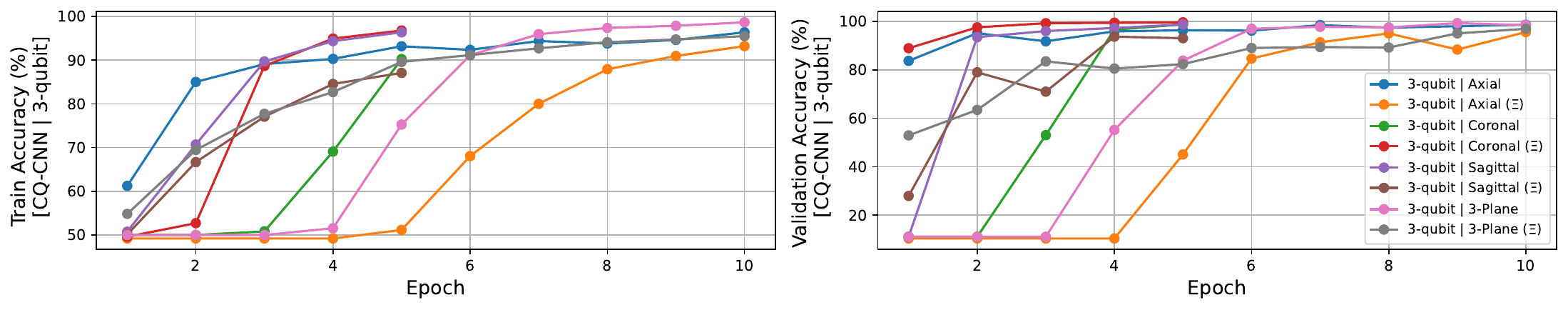}
    \end{subfigure}
    \caption{The graphs present the training and validation accuracy curves for classification models across different MRI planes (axial, coronal, sagittal, and 3-plane) and model configurations (classical, 2-qubit, and 3-qubit), with and without skull-stripping, over several epochs. The classical CNN (top row) shows a steady, step-by-step improvement in accuracy with each epoch. In contrast, the CQ-CNN models (middle and bottom rows) exhibit slow convergence during the initial phase of training but rapidly achieve high accuracy after a few more epochs.}
    \label{fig:train-val}
\end{figure*}

\subsubsection*{Classical-quantum convergence analysis}
In our experiments with CQ-CNN models, we discovered a few repetitive patterns during training, particularly in the initial phases. The MRI images from both the nondementia (healthy) and moderate dementia (Alzheimer’s) classes are often highly similar, making it difficult for the model to discern subtle differences between the two. While quantum models are theoretically well-suited for handling high-dimensional data and capturing intricate patterns, they face practical limitations when dealing with subtle class distinctions, as seen in AD classification datasets. The primary issue arises from the quantum component of the architecture, which, despite its refined design, struggles with convergence in the early stages of training, as shown in the middle and bottom rows of Figure \ref{fig:train-val}. In classical CNN models, we usually address this issue by increasing the number of parameters, enabling the model to better capture relevant features from the training data. However, when this approach is applied to quantum models by increasing the number of qubits, convergence failure worsens instead of improving.

One major reason for this instability is the inability of quantum gates to produce well-defined gradients. Quantum circuits, particularly those that use feature maps and ansatz, often result in poor gradient flow during optimization, especially when dealing with datasets in which images within the classes have fewer discriminative features. This can cause gradients to vanish or explode, making it difficult for the optimizer to adjust the quantum weights effectively. The classical CNN component, responsible for gradient-based optimization, functions well in its domain, but its optimization strategies often fail to translate smoothly to the quantum part of the model. This disconnect leads to poor convergence, particularly in the initial phase of training. As a result, CQ-CNN models often require multiple re-runs of experiments before achieving satisfactory performance.

That said, when properly converged, CQ-CNN models perform well, requiring fewer epochs than classical models to reach their potential accuracy. For example, when comparing the classical model with the 2-qubit model trained on coronal images (Figure \ref{fig:train-val}, top row: classical model, middle row: quantum model, green line), the classical model requires five epochs to exceed 95\% accuracy, whereas the quantum model achieves this in just two epochs. This demonstrates that the quantum advantage remains evident in our experiments, despite being overshadowed by convergence failures.

\subsection{Ablation study}
\textbf{Gradient optimization algorithm tuning: } To determine which gradient optimization algorithm works best for our CQ-CNN models, we experimented with several options. Adam was the only optimizer that successfully enabled our models to converge, so we used it in all our experiments. In contrast, all other optimizers we tested, including SGD, L-BFGS, RMSprop, and Adagrad, failed to do so. Their failure can be attributed to the highly non-convex loss landscapes and gradient instability of quantum neural networks. For instance, SGD, which relies on small, incremental updates, becomes unreliable in quantum architectures due to gradient noise and non-smooth loss surfaces. L-BFGS, a second-order optimization method, assumes well-behaved loss functions, an assumption that rarely holds in hybrid quantum-classical models, leading to poor convergence. RMSprop and Adagrad, which adjust learning rates based on past gradients, struggle due to quantum parameter sensitivity, often resulting in excessively small updates that limit meaningful learning progress. In contrast, Adam’s momentum-based adaptive learning strategy helps stabilize erratic gradients, making it more resilient in CQ-CNN training. Despite initial struggles, Adam eventually adapted to optimize the quantum parameters of the PQC in CQ-CNN models, enabling the model to learn effectively in the later stages of training.

\begin{table*}[ht!]
    \centering
    \caption{Comparison of classical models from recent literature and our proposed classical-quantum models for AD detection, detailing key attributes such as the dataset, number of classes, model type, segmentation usage, accuracy, number of parameters, and model size.}
    \label{tab:comparison}
    \renewcommand{\arraystretch}{1.3}
    \resizebox{1.0\textwidth}{!}{
    \begin{tabular}{
        l@{\hspace{10pt}}
        l@{\hspace{8pt}}
        l@{\hspace{8pt}}
        c@{\hspace{8pt}}
        l@{\hspace{8pt}}
        c@{\hspace{8pt}}
        l@{\hspace{10pt}}
        l@{\hspace{8pt}}
        l}
        \toprule
        Method & Year & Dataset & Class & Model Type & Segmentation & Accuracy & Parameters & Size (MB) \\
        \midrule
        AlexNet \cite{nawaz2021deep} & 2020 & OASIS & 2+ & Classical & \(\times\) & 0.9285 & 60M & 227.0 \\
        ResNet-50 \cite{sun2021improved} & 2021 & ADNI & 2+ & Classical & \checkmark & 0.971 \(\pm\) 0.016 & 25.6M & 98.0 \\
        DenseCNN2 \cite{katabathula2021predict} & 2021 & ADNI & 2 & Classical & \(\times\) & 0.9252 & 7.1M & 28.4 \\
        2D-M\(^2\)IC \cite{helaly2022deep} & 2022 & ADNI & 2 & Classical & \(\times\) & 0.9711 & 10.3M & 39.5 \\
        3D-M\(^2\)IC \cite{helaly2022deep} & 2022 & ADNI & 2 & Classical & \(\times\) & 0.9736 & 18.2M & 69.7 \\
        Ensemble \cite{jenber2024deep} & 2024 & ADNI & 2 & Classical-Quantum & \(\times\) & 0.9989 & 27.5M & 105.2 \\
        ResNet-101 \cite{ghaffari2022deep} & 2022 & ADNI/OASIS & 2 & Classical & \checkmark & 0.9078 & 44.5M & 171.0 \\
        Xception \cite{ghaffari2022deep} & 2022 & ADNI/OASIS & 2 & Classical & \checkmark & 0.8438 & 22.9M & 88.0 \\
        Inception-v3 \cite{ghaffari2022deep}  & 2022 & ADNI/OASIS & 2 & Classical & \checkmark & 0.9375 & 23.9M & 92.0 \\
        2D-CNN\cite{castellano2024automated} & 2024 & OASIS-3 & 2 & Classical & \checkmark & 0.7793 & 4.3M & 16.5 \\
        3D-CNN \cite{castellano2024automated} & 2024 & OASIS-3 & 2 & Classical & \checkmark & 0.9167 & 5.8M & 22.3 \\
        \midrule
        \textbf{\(\bm{\alpha_8}\)-2-qubit} & \textbf{2025} & \textbf{OASIS-2} & \textbf{2} & \textbf{Classical-Quantum} & \checkmark & \textbf{0.9350 \(\pm\) 0.0390} & \textbf{13K + 2 qubits} & \textbf{0.48} \\ 
        \textbf{\(\bm{\beta_8}\)-3-qubit} & \textbf{2025} & \textbf{OASIS-2} & \textbf{2} & \textbf{Classical-Quantum} & \checkmark & \textbf{0.9750 \(\pm\) 0.0076} & \textbf{13K + 3 qubits} & \textbf{0.48} \\ 
        \bottomrule
    \end{tabular}
    }
\end{table*}

\textbf{Classical parameters tuning: } We also experimented with increasing the classical parameters of the neural network by adding larger convolutional filters. However, the issue persisted, leading us to conclude that effective training of quantum models cannot be achieved simply by adding more qubits, increasing parameters, or making the architecture more complex. Instead, the focus should be on refining the gradient optimization process.

\subsection{Comparative and computational analysis with existing methods}
Table \ref{tab:comparison} compares classical CNN models with our proposed classical-quantum CNN models for AD detection based on performance and computational factors. The key insights from this comparison are discussed below.  

First notable distinction is the computational setup for training. Classical CNN models are typically trained on GPUs, benefiting from well-established deep learning frameworks optimized for GPU acceleration. In contrast, classical-quantum models, including ours, rely on CPUs, as no efficient mechanism currently exists to fully utilize GPU computation for training such networks. Even when we experimented with training the classical portion on a GPU while keeping the quantum portion on a CPU, we observed no significant improvement in training time. Consequently, classical-quantum models remain limited to CPU-based simulations, making training on large datasets significantly more time-consuming than with classical models. Despite this limitation, we successfully trained our model on over a thousand images while maintaining rigorous experimental standards.  

Another key observation is the role of brain tissue segmentation in AD classification. Many classical models, such as ResNet-50 (used by Sun \textit{et al.} (2021) \cite{sun2021improved}), ResNet-101, Xception, and Inception-v3 (utilized by Ghaffari \textit{et al.} (2022) \cite{ghaffari2022deep}), as well as custom 2D and 3D CNNs (developed by Castellano \textit{et al.} (2024) \cite{castellano2024automated}), incorporate skull-stripping or brain tissue segmentation before classification. Ghaffari \textit{et al.} used a U-Net for segmentation, while Castellano \textit{et al.} applied the Otsu threshold method. In our study, we introduce SkullNet, a U-Net-based architecture designed specifically for brain tissue segmentation across all three anatomical MRI planes. SkullNet processes 128×128×1 input images and generates segmentation masks of the same resolution (as depicted in Figure \ref{fig:u-net}). To support future research, we have publicly released SkullNet, enabling other researchers to bypass the need for custom segmentation model training.  

Regarding classification performance and computational complexity, we highlight several key observations: \textbf{\textit{(i)}} Our \(\beta_8\)-3-qubit model achieves an accuracy of 0.9750 on the OASIS-2 dataset, which is on par with, and in some cases even better than, the classical SOTA models. \textbf{\textit{(ii)}} While AlexNet \cite{nawaz2021deep} achieved the highest previous OASIS accuracy of 0.9285, it employs approximately 60 million parameters (227 MB). In stark contrast, our model uses merely 13K parameters (0.48 MB), a reduction of over 99.99\%. This underlines the substantial potential of quantum computing to deliver superior performance with significantly fewer parameters. \textbf{\textit{(iii)}} The 3D-CNN (proposed by Castellano \textit{et al.} 2024 \cite{castellano2024automated}) reaches 0.9167 accuracy on OASIS-3 using 5.8 million parameters. Our model outperforms this with only 0.24\% of the parameter count, reinforcing the scalability of our approach. \textbf{\textit{(iv)}} Models such as the Ensemble by Jenber \textit{et al.} (2024) \cite{jenber2024deep} and 3D-M\(^2\)IC by Helaly \textit{et al.} (2022) \cite{helaly2022deep} report competitive accuracies (0.9989 and 0.9736, respectively) on ADNI. However, they utilize substantially larger parameter counts (27.5M and 18.2M, respectively) and are trained on significantly larger datasets. They used 38,400 images (256×256), whereas our model achieved higher performance with only ~5,900 images (128×128). This also signifies the efficiency of our method in data-limited clinical scenarios. \textbf{\textit{(v)}} Jenber \textit{et al.} (2024) \cite{jenber2024deep}, although holding SOTA on the ADNI dataset, comes with concerning issues. They extract features from an ensemble of CNNs and feed the concatenated features into a Quantum Support Vector Machine (QSVM), where the quantum circuit only computes kernel values for classical SVM classification. This hybrid design lacks end-to-end quantum optimization. In contrast, our model optimizes the trainable quantum parameters through an end-to-end pipeline, using parameter-shift gradient calculations in a fully differentiable manner. \textbf{\textit{(vi)}} Many prior methods (e.g., \cite{jenber2024deep, helaly2022deep}) employed rotations or reflections that can distort critical anatomical features for AD analysis. Our model uses a diffusion-based strategy to generate synthetic images of the minority class while preserving anatomical orientation, thereby minimizing the risk of introducing artifacts that could undermine clinical reliability. \textbf{\textit{(vii)}} Our approach achieves a significant reduction in model size (0.48 MB compared to tens or hundreds of MB for traditional models), offering considerable practical benefits for deployment in clinical settings with limited memory or computational resources. Consequently, our method not only delivers SOTA performance on OASIS data but also does so with exceptional parameter efficiency, highlighting a distinct quantum advantage in space complexity for AD classification.

\section{Discussion}
\label{section:5}
The findings from our experiments with CQ-CNN models for AD detection can be divided into several parts. In the first part, we investigate the potential challenges of embedding a PQC into a CNN during training. We experiment with various architectural changes, such as increasing the number of qubits, adjusting classical parameters, modifying the size of the dataset, and altering the number of classes. Through numerous trials and errors, we identify that the primary factor contributing to the models' initial low convergence is the high similarity between images from different classes, so much so that even the human eye struggles to distinguish between them.

To elaborate on this factor, while our primary focus is binary classification, we also experimented with a multi-class classification setup, attempting to distinguish between four closely related AD classes. In this case, the convergence issue became significantly worse, with the model almost failing to converge. Even when it did converge, the process was extremely slow. When we reverted to binary classification, the situation improved. Interestingly, when applying a classical CNN model to the same four-class classification task, we did not encounter the same problem. This suggests that the high similarity of images and the increased classification complexity negatively affect the convergence of the quantum model.

In the second part, we investigate the potential causes of low convergence within the architecture. We found that the underlying issue appears to stem from the gradient, which is responsible for updating the model's weights. If the gradient gets stuck in a local optimum, it incorrectly assumes that it has reached the optimal solution, preventing further improvements. As a result, the quantum weights remain unchanged, leading to poor convergence as the task complexity increases. This suggests that similar convergence issues could occur in other medical imaging classification tasks using CQ-CNN-like architectures, though additional experiments are necessary for a definitive conclusion. However, our findings also indicate that when the quantum model successfully converges, it demonstrates a potential quantum advantage, even in distinguishing closely related classes, such as moderate dementia and non-dementia, in AD detection.

In the final section, we conducted an extensive comparative analysis with existing models, revealing a significant advantage in computational efficiency. The drastic reduction in parameter count (over 99.99\% compared to traditional CNNs like AlexNet), while achieving better accuracy, demonstrates that quantum models have the potential to replace classical models in medical image analysis, making high-performance diagnostic tools more accessible in resource-constrained clinical settings. The improved computational efficiency, combined with our diffusion-based data generation approach that preserves anatomical integrity, suggests that CQ-CNN architectures could become particularly valuable in specialized medical domains where data is limited but diagnostic accuracy is critical.

\section{Conclusion}
\label{section:6}
The automatic detection of Alzheimer’s disease (AD) from MRI images using classical CNN models is a widely researched area in medical image analysis. However, with the continuous emergence of new technologies, it is essential to explore the next computational frontier: quantum computing. Although relatively new, quantum computing has already been integrated with classical machine learning (CML), as seen in parameterized quantum circuit (PQC)-embedded CNN networks. In this research, we propose CQ-CNN, a hybrid classical-quantum convolutional neural network architecture for binary image classification. To ensure clinically reliable results, we take several major steps before training our classifiers. First, we develop a simple framework to convert 3D volumetric MRI data into 2D slices, which is essential for the classification task. Next, we address a common challenge in both CML and QML: class imbalance. To solve this, we train a custom probabilistic diffusion model to generate synthetic images of the minority class. In addition, to ensure our models focus on features relevant to AD detection, we train a multi-view segmentation model named SkullNet, which extracts brain tissue while removing irrelevant structures like the skull and surrounding areas. Using the 3D-to-2D conversion framework, diffusion model, and segmentation model, we generate several variations of our classification dataset and train our classifiers. 

Our experiments reveal a significant limitation in the current hybrid classical-quantum architecture for image classification. We found that when images within a class are highly similar, the quantum model struggles to converge due to gradient failure, resulting in minimal weight updates and the model getting stuck during optimization. We believe this issue might also affect other medical imaging datasets and propose it as a direction for future research. However, when the model does converge, we observe evidence of quantum advantage, as quantum models require significantly fewer epochs to achieve comparable accuracy compared to classical models. Our \(\beta_8\)-3-qubit model established a new SOTA benchmark on the OASIS-2 dataset with an accuracy of 0.9750, while requiring only 13K parameters (0.48 MB), which is orders of magnitude fewer than current methods. This parameter efficiency demonstrates that quantum approaches can potentially deliver improved diagnostic performance with substantially reduced computational requirements, which is especially valuable for deployment in clinical environments. These findings suggest that, with further improvements in quantum optimization techniques, hybrid classical-quantum architectures could become valuable tools for medical imaging tasks such as AD detection.

\section*{Data availability}
The datasets used in this research are publicly accessible at the following links: 
\href{http://preprocessed-connectomes-project.org/NFB_skullstripped}{NFBS dataset} and 
\href{https://sites.wustl.edu/oasisbrains/home/oasis-2}{OASIS-2 dataset}.

\section*{CRediT authorship contribution statement}
\textbf{M. Islam:} Conceptualization, Methodology, Software, Validation, Formal analysis, Investigation, Resources, Data Curation, Writing - Original Draft, Writing - Review \& Editing, Visualization, Project administration.
\textbf{M. Hasan:} Conceptualization, Formal analysis, Validation, Writing - Original Draft, Writing - Review \& Editing, Supervision, Project administration. \textbf{M.R.C. Mahdy:} Conceptualization, Validation, Writing - Review \& Editing, Supervision.

\section*{Declaration of competing interest}
The authors declare that they have no known competing financial interests or personal relationships that could have appeared to influence the work reported in this paper.




\end{document}